\documentclass[12pt]{article}
\usepackage{amsfonts,amsmath,amssymb,epsf,framed}
\usepackage{amsmath,braket}
\usepackage{graphicx,hyperref,color,comment}
\usepackage{cite}
\usepackage{dsfont} 
\usepackage{bm} 
\usepackage{appendix}

\usepackage{subfigure}

\edef\restoreparindent{\parindent=\the\parindent\relax}
\usepackage{parskip}
\restoreparindent

\hypersetup{
    bookmarks=true,         
    unicode=false,          
    pdftoolbar=true,        
    pdfmenubar=true,        
    linktocpage=true,       
    pdffitwindow=false,     
    pdfstartview={FitH},    
    pdfnewwindow=true,      
    colorlinks=true,       
    linkcolor=blue,          
    citecolor=blue,        
    filecolor=blue,      
    urlcolor=blue           
}
\numberwithin{equation}{section}									

\newcommand{\sgn}{\textrm{sgn}}
\newcommand{\la}{\langle}
\newcommand{\ra}{\rangle}

\newcommand{\eff}{\textrm{eff}}
\newcommand{\tree}{\textrm{tree}}
\newcommand{\oneloop}{\textrm{one-loop}}

\let\a=\alpha \let\b=\beta  \let\d=\delta  \let\ve=\varepsilon \let\g=\gamma  \let\k=\kappa \let\l=\lambda \let\m=\mu \let\n=\nu
  \let\r=\rho \let\s=\sigma \let\t=\tau \let\th=\theta  \let\vp=\varphi   
  \let\G=\Gamma \let\L=\Lambda   \let\S=\Sigma \let\Th=\Theta   
\def\nn{\nonumber}
\def\inf{\infty}

\DeclareMathOperator{\Tr}{Tr}

\begin{document}

\begin{titlepage}
\thispagestyle{empty}
\rightline{\tt OU-HET-1324}
\vspace*{1cm}

\begin{center}
\noindent{{\Large \textbf{Open-Closed-Open Triality\\[14pt] for 1/2 BPS Three-Point Functions}}}\\
\vspace{1.5cm}

Keisuke Konosu$^a$ \, and \, Kenta Suzuki$^b$
\vspace{1cm}\\

{\it ${}^a$Department of Physics, The University of Osaka, \\[2pt] Machikaneyama-Cho 1-1, Toyonaka 560-0043, Japan}, \\[8pt]
{\it ${}^b$Graduate School of Arts and Sciences, University of Tokyo, \\[2pt] Komaba, Meguro-ku, Tokyo 153-8902, Japan}

\vskip 6em
\end{center}

\begin{abstract}
Open-closed-open triality relates two distinct open string descriptions through a common closed string theory.
In this paper, we develop an open-closed-open triality for protected half-BPS three-point functions in four-dimensional $\mathcal{N}=4$ super Yang-Mills theory.
Starting from determinant insertions representing giant gravitons, we formulate a V-type open-string Gaussian three-matrix model and,
through a color-flavor transformation, derive an F-type open-string model of three bifundamental matrices forming a triangular quiver.
Its closed quiver walks include oriented cubic cycles, a feature absent from the bipartite two-point case.
We test the correspondence by deriving exact finite-$N$ single-giant polynomials, recovering the known F-type large-$N$ saddles,
and proving the equivalence of the V- and F-type character expansions via an explicit bijection of their combinatorial data.
At the closed string corner, connected three-matrix contractions determine three-colored integer Strebel surfaces.
We argue that the corresponding target-space construction is naturally described by a barycentrically subdivided Belyi map or a colored Hurwitz constellation.

\end{abstract}

\end{titlepage}

\newpage

\tableofcontents

\section{Introduction}
\label{sec:introduction}

Gauge/gravity duality or holography principle \cite{tHooft:1993dmi, Susskind:1994vu} gives a non-perturbative relation between gauge theories and quantum gravity.
However, the microscopic reorganization of degrees of freedom from both sides remains difficult to exhibit explicitly.
In the large-$N$ expansion, double-line diagrams carry a genus and therefore have the topology expected from string perturbation theory \cite{tHooft:1973alw}. 
Turning this topological observation into a concrete worldsheet construction requires a prescription that assigns moduli, and not only a genus, to every ribbon graph.

The AdS/CFT correspondence \cite{Maldacena:1997re} provides the most accurately formulated duality of this kind.
The correspondence identifies a quantum theory of gravity on an asymptotically AdS spacetime with a conformal field theory without gravity on its boundary. 
Its canonical example is the duality between type IIB superstring theory on $\mathrm{AdS}_{5}\times S^{5}$, supported by $N$ units of self-dual
Ramond--Ramond five-form flux, and four-dimensional $\mathcal{N}=4$ super Yang--Mills (SYM) theory with $SU(N)$ gauge symmetry \cite{Aharony:1999ti,DHoker:2002nbb,Ammon_Erdmenger_2015}.
The $U(N)$ theory used below differs only by a decoupled center-of-mass $U(1)$ sector. 
The Gubser--Klebanov--Polyakov--Witten prescription \cite{Gubser:1998bc,Witten:1998qj} equates the bulk string partition function with the generating functional of boundary correlators, thereby mapping bulk fields and states to local gauge-invariant operators.

The half-BPS sector of four-dimensional $\mathcal{N}=4$ SYM theory is a particularly useful setting for the problem. 
The field content of $\mathcal{N}=4$ SYM includes six real adjoint scalars transforming under the $SO(6)\simeq SU(4)$ R-symmetry. 
Choosing a complex combination $Z$ of two scalars, one obtains half-BPS chiral primary operators with protected dimension $\Delta=J$, where $J$ is a $U(1)$ R-charge. 
Its two- and three-point functions are protected and may be computed in the free theory \cite{Lee:1998bxa,DHoker:1998vkc,Baggio:2012rr}, while they still retain non-trivial finite-$N$ combinatorics. 
Operators whose charges scale with $N$ probe the non-planar regime and describe D-branes in the bulk. 
In particular, determinant and Schur-polynomial operators furnish gauge-theory descriptions of giant gravitons \cite{McGreevy:2000cw,Grisaru:2000zn,Hashimoto:2000zp,Balasubramanian:2001nh,Corley:2001zk}.
They therefore provide a natural arena in which both open and closed strings can be seen within the same protected observable.

A direct route from free correlators to worldsheets was developed in \cite{Gopakumar:2003ns,Gopakumar:2004qb,Gopakumar:2005fx}. 
After reducing a Feynman diagram to a metrized skeleton, its dual ribbon graph is identified with the critical graph of a Strebel differential.
The Strebel correspondence then associates the diagram with a point in the decorated moduli space of punctured Riemann surfaces \cite{Strebel:1984qd,Kontsevich:1992ti,Mulase:1998un,Razamat:2008zr}. 
When the edge lengths are integers, this construction is closely related to dessins d'enfants, Belyi maps, and permutation descriptions of branched covers~\cite{Grothendieck:1984edp,Lando:2004gsa,MR534593}. 
Thus Feynman diagrams label discrete points in moduli space rather than merely furnishing a lattice discretization of the worldsheet.

Open--closed--open triality sharpens this picture: two brane stacks can yield, by integrating out either stack, distinct open string descriptions encoding the same closed background.
Diagrammatically, closed-string insertions occur at vertices in the V-type description and at faces in the F-type description,
so their ribbon expansions are related by graph duality \cite{Gopakumar:2010bd,Gopakumar:2022djw}.\footnote{See also the recent progress of this topic~\cite{Mazenc:2026yrm}.} 
The paradigmatic solvable example is the $(2,1)$ minimal string, or pure topological gravity \cite{Witten:1990hr,DiFrancesco:1993nw}. 
Its V-type corner is the double-scaling limit at the spectral edge of the Hermitian Gaussian matrix integral, whose eigenvalue/ZZ-brane degrees of freedom build the closed background; determinant insertions describe FZZT probes.
Rewriting their characteristic-polynomial correlators by integrating fields in and out, and then taking the same double-scaling limit, gives the Kontsevich matrix Airy model \cite{Brezin:1999jv,Brezin:2007xi,Maldacena:2004sn}. 
This is cubic open-string field theory on FZZT branes, with Miwa variables generating closed gravitational descendants and intersection numbers \cite{Kontsevich:1992ti,Gaiotto:2003yb,Hashimoto:2005fr}. 
Thus, the Gaussian and Kontsevich integrals are distinct open descriptions of one closed minimal string, with vertices and faces exchanged.

The same organization appears in the half-BPS sector of $\mathcal{N}=4$ SYM.
Determinant operators representing giant gravitons are inserted in an $N\times N$ Gaussian two-matrix integral: this is the V-type open description, whose color ribbon graphs generate the common closed worldsheets. 
Integrating out the color matrices instead yields a lower-rank flavor matrix integral for open strings between giant-graviton stacks, the F-type corner \cite{Brown:2010af,Jiang:2019xdz,Gopakumar:2024jfq}. 
For two-point functions the graphs are bipartite.  At three points the protected correlator requires three matrices~\cite{Kazakov:2024ald,Anempodistov:2025maj},
triangular Wick cycles and odd-length index-loop faces are possible, so the extension is nontrivial; this work develops the resulting three-node triality.

In this paper we construct a three-point generalization of the open-closed-open framework. 
Starting with products of determinant operators at three spacetime points, we obtain a V-type three-matrix model with three colors of trace vertices. 
We then rewrite the determinants using auxiliary fermions, integrate out the $N\times N$ color matrices,
and apply Hubbard--Stratonovich transformations of the type used in matrix integrals with characteristic polynomials and external sources \cite{Maldacena:2004sn,Brezin:2016eax}. 
The resulting F-type theory contains three bifundamental matrices forming the edges of a triangular three-node quiver. 
Its interactions enumerate closed quiver walks; in particular, cubic oriented triangles are the new feature absent from the two-point model.

We test the dual descriptions in several complementary ways. 
In the single-giant sector we compare the F-type large-$N$ saddle analysis with exact finite-$N$ V-type polynomials for both two- and three-point functions. 
We also develop character expansions of the two descriptions using Schur--Weyl duality and symmetric-group representation theory,
and show how their representation data and contraction channels are identified. 
Restricting the character expansion to one determinant at each point reproduces the single-giant finite-$N$ polynomial. 
At the closed-string corner, every connected V-type contraction defines a three-colored ribbon graph and hence an integer Strebel surface. 
The worldsheet construction therefore extends directly, whereas the target-space description naturally requires either a barycentrically subdivided Belyi map or a colored Hurwitz/constellation description.

The paper is organized as follows.  In section~\ref{sec:triality} we review the two-point V/F-type matrix models and then derive a new open-closed-open triality for three-point functions. 
Section~\ref{sec:single} studies the single-giant sector, and section~\ref{sec:character} develops the character expansions and their comparison. 
Section~\ref{sec:string} describes the worldsheet and target-space constructions, and then we conclude in section~\ref{sec:conclusions}.

\section{Open-Closed-Open Triality}
\label{sec:triality}
In this section, we discuss the notion of the open-closed-open triality~\cite{Gopakumar:2010bd, Gopakumar:2022djw, Gopakumar:2024jfq}. In section~\ref{sec:mat-reduction}, we consider some specific classes of the correlation functions of $1/2$-BPS operators and discuss their reductions to matrix integrals following~\cite{Brown:2010af,Gopakumar:2024jfq}. 
In section~\ref{sec:twopt-MatrixDuality}, we review the duality between two open string descriptions of the two-point functions of $1/2$-BPS operators~\cite{Brown:2010af,Gopakumar:2024jfq}. 
In section~\ref{sec:threept-MatrixDuality}, we derive a new dual open string descriptions of the matrix integral derived from the three-point functions of $1/2$-BPS operators. 

\subsection{Reduction to matrix integrals}
\label{sec:mat-reduction}
Let us consider $n$-point functions of the $1/2$-BPS operators in $\mathcal{N}=4$ SYM theory with $U(N)$ gauge group in four dimensions of Euclidean signature. We focus on correlation functions of $1/2$-BPS operators $\mathcal{O}_{i}$ inserted at a position $w_{i}$ involving linear combinations of two out of the six scalars $\Phi^{I}$ with $I=1,2,\ldots,6$. It is defined by
\begin{align}
    \left\langle\,\prod_{i=1}^{n}\mathcal{O}_{i}(Y_{i}\cdot\Phi(w_{i}))\,\right\rangle_{\mathcal{N}=4}
    =\frac{1}{Z_{\mathcal{N}=4}}\int(\mathrm{fields})\,e^{-\frac{N}{\lambda}S_{\mathcal{N}=4}}\,\prod_{i=1}^{n}\mathcal{O}_{i}(Y_{i}\cdot\Phi(w_{i}))\,,\label{eq:original-correlator}
\end{align}
where $S_{\mathcal{N}=4}$ is the action of $\mathcal{N}=4$ SYM theory with the ’t Hooft coupling $\lambda$, and $Z_{\mathcal{N}=4}$ is its partition function.
We also introduced complex null polarization vectors $Y_{i}^I$ which satisfy $Y_i \cdot Y_i = 0$.
The symbol $\cdot$ represents the inner product $Y_{i}\cdot\Phi=Y_{i}^{I}\Phi^{J}\delta_{IJ}$.

When we focus on $n=2$ or $n=3$ cases, the correlation functions~\eqref{eq:original-correlator} are protected~\cite{Lee:1998bxa,DHoker:1998vkc,Baggio:2012rr} and the interaction terms can be neglected. 
Since the correlation functions~\eqref{eq:original-correlator} only depend on the scalar fields $\Phi^I$, they are reduced to
\begin{align}
    \left\langle\,\prod_{i=1}^{n}\mathcal{O}_{i}(Y_{i}\cdot\Phi(w_{i}))\,\right\rangle_{\mathcal{N}=4}=\frac{1}{Z_{\mathrm{scalar}}}\int D\Phi^I\,e^{\frac{N}{2\lambda}\int d^{4}w\,\Tr(\Phi^{I}\Box\Phi^{I})}\prod_{i=1}^{n}\mathcal{O}_{i}(Y_{i}\cdot\Phi(w_{i}))\,,
\end{align}
where $Z_{\mathrm{scalar}}$ is the partition function of scalar fields in $\mathcal{N}=4$ SYM. The scalar propagator is given by
\begin{align}
    \langle\,\Phi^I(w_{1})\,\Phi^J(w_{2})\,\rangle=\frac{\lambda\,\delta^{IJ}}{4\pi^{2}Nw_{12}^{2}}\,,\label{eq:scalar-propagator}
\end{align}
where $w_{ij}=w_{i}-w_{j}$, and we have
\begin{align}
     \langle\,Y_{1}\cdot\Phi(w_{1})\,Y_{2}\cdot\Phi(w_{2})\,\rangle=\frac{\lambda\,Y_{1}\cdot Y_{2}}{4\pi^{2}Nw_{12}^{2}}\,.
\end{align}
Since the action is free, we can extract the dependence of coordinates and the correlators are reduced to the matrix integrals.

Let us first consider $n=2$ case. 
The correlation functions~\eqref{eq:original-correlator} are reduced to\footnote{We note that the integration contour for $K$ is rotated to imaginary axis. See footnote 42 and Appendix B of~\cite{Gopakumar:2024jfq} for details.} 
\begin{align} 
    \left\langle\,\prod_{i=1}^{2}\mathcal{O}_{i}(Y_{i}\cdot\Phi(w_{i}))\,\right\rangle_{\mathcal{N}=4}=\frac{1}{Z_{2\mathrm{pt}}}\int dK dM\,e^{-\frac{N}{g}S[K,M]}\,\mathcal{O}_{1}(K)\,\mathcal{O}_{2}(M)\,.\label{eq:2pt-Vtype-ini}
\end{align}
where $K$ and $M$ are $N \times N$ Hermitian matrices with
\begin{align}
    g=\frac{Y_{1}\cdot Y_{2}}{4\pi^{2}w_{12}^{2}}\lambda\,,
\end{align}
and
\begin{align}
    S[K,M]=\Tr(KM)\,.
\end{align}
The matrices $K$ and $M$ are introduced to reproduce the two point functions~\eqref{eq:scalar-propagator} by formally replacing $Y_{1}\cdot\Phi(w_{1}),\,Y_{2}\cdot\Phi(w_{2})$ with $K$ and $M$, respectively.
The normalization factor $Z_{2\mathrm{pt}}$ is defined such that (\ref{eq:2pt-Vtype-ini}) is unity for identity operators.\footnote{We comment that the matrix integral we consider in this paper does not converge. They should be understood as the formal matrix integral, which reproduce the original $\mathcal{N}=4$ SYM result by Wick contractions.}

Similarly, the three-point functions are reduced to~\cite{Kazakov:2024ald,Anempodistov:2025maj} 
\begin{align} 
    \left\langle\,\prod_{i=1}^{3}\mathcal{O}_{i}(Y_{i}\cdot\Phi(w_{i}))\,\right\rangle_{\mathcal{N}=4}=\frac{1}{Z_{3\mathrm{pt}}}\int \prod_{\alpha=1}^{3} dM_{\alpha}\,e^{-\frac{2\pi^{2}N}{\lambda}S[M_{1},M_{2},M_{3}]}\prod_{i=1}^{3}\mathcal{O}_{i}(M_{i})\,,
\end{align}
where
\begin{align}
    S[M_{1}, M_{2}, M_{3}] = \Tr
    \begin{pmatrix}
        M_{1} & M_{2} & M_{3} 
    \end{pmatrix}
        \begin{pmatrix}
            0 & p_{12} & p_{13} \\
            p_{12} & 0 & p_{23} \\
            p_{13} & p_{23} & 0
        \end{pmatrix}^{-1}
    \begin{pmatrix}
        M_{1} \\
        M_{2} \\
        M_{3}
    \end{pmatrix}
\end{align}
and
\begin{align}
    p_{\alpha\beta}=\frac{Y_{\alpha}\cdot Y_{\beta}}{w_{\alpha\beta}^{2}}\,,
\end{align}
with $\a,\b=1,2,3$ and $\a\neq \b$. 
Since
\begin{align}
    \begin{pmatrix}
            0 & p_{12} & p_{13} \\
            p_{12} & 0 & p_{23} \\
            p_{13} & p_{23} & 0
    \end{pmatrix}^{-1}
    =\frac{1}{2}
    \begin{pmatrix}
           -\frac{p_{23}}{p_{12}p_{13}}  & \frac{1}{p_{12}} & \frac{1}{p_{13}} \\
            \frac{1}{p_{12}} & -\frac{p_{13}}{p_{12}p_{23}} & \frac{1}{p_{23}} \\
            \frac{1}{p_{13}} & \frac{1}{p_{23}} & -\frac{p_{12}}{p_{23}p_{13}}
    \end{pmatrix}\,,
\end{align}
the reduced theory is a three Hermitian matrix model with the matrix size $N\times N$, whose action is given by
\begin{align}
    S[M_{1},M_{2},M_{3}]
        =-\frac{1}{2}\sum_{\alpha=1}^{3}\frac{1}{\kappa_{\alpha}^{2}}\Tr M_{\alpha}^{2}+\frac{1}{2}\sum_{\substack{\alpha,\beta=1\\\alpha\neq\beta}}^{3}\frac{1}{\kappa_{\alpha}\kappa_{\beta}}\Tr(M_{\alpha}M_{\beta})\,.
\end{align}
where
\begin{align}
    \kappa_{1}=\left(\frac{p_{12}p_{13}}{p_{23}}\right)^{\frac{1}{2}}\,,\qquad \kappa_{2}=\left(\frac{p_{12}p_{23}}{p_{13}}\right)^{\frac{1}{2}}\,,\qquad\kappa_{3}=\left(\frac{p_{13}p_{23}}{p_{12}}\right)^{\frac{1}{2}}\,.
\end{align}
The Hermitian matrices $M_{i}$ are introduced to reproduce the two point functions~\eqref{eq:scalar-propagator} by formally replacing $Y_{i}\cdot\Phi(w_{i})$ to $M_{i}$,
For notational simplicity, we put
\begin{align}
    g=\frac{\lambda}{2\pi^{2}}
\end{align}
and consider
\begin{align} 
    \left\langle\,\prod_{i=1}^{3}\mathcal{O}_{i}(Y_{i}\cdot\Phi(w_{i}))\,\right\rangle_{\mathcal{N}=4}=\frac{1}{Z_{\mathrm{3pt}}}\int \prod_{\alpha=1}^{3} dM_{\alpha}\,e^{-\frac{N}{g}S[M_{1},M_{2},M_{3}]}\prod_{i=1}^{3}\mathcal{O}_{i}(M_{i})
\end{align}
with the partition function $Z_{\mathrm{3pt}}$.

\subsection{Two-point functions}
\label{sec:twopt-MatrixDuality}
In this subsection, we review the duality between two open string descriptions for two-point functions of~\cite{Gopakumar:2024jfq, Brown:2010af}. Let us start from the \textit{V(ertex)-Type} open dual. We begin with the matrix integral, which is given by
\begin{equation}
    \begin{split}
        Z_{\mathrm{2det}} \, = \, Z_{2V} \, &:= \, \frac{1}{(2\pi g)^{N^{2}}}\left(\frac{\det_{N}(NY)}{\det_{Q}(X)\det_{R}(V)}\right)^{N}\\
        &\quad\times\int dK dM_{N\times N}\,e^{-\frac{N}{g}\Tr\sqrt{Y}K\sqrt{Y}M}\prod_{a=1}^{Q}\det(x_{a}-M)\prod_{a'=1}^{R}\det(v_{a'}-K)\,,
    \end{split}
\end{equation}
where the diagonal source matrices are 
\begin{align}
    Y_{N\times N}=\mathrm{diag}(y_{1},\ldots,y_{N})\,,\quad
    X_{Q\times Q}=\mathrm{diag}(x_{1},\ldots,x_{Q})\,,\quad
    V_{R\times R}=\mathrm{diag}(v_{1},\ldots,v_{R})\,.
\end{align}
This matrix integral can be obtained with the field redefinition and the specific operator insertions from~\eqref{eq:2pt-Vtype-ini}.
These determinant operators are interpreted as giant gravitons in the bulk of AdS/CFT \cite{Balasubramanian:2001nh, Corley:2001zk}.
Since $\det(A)=e^{\Tr\log(A)}$ with a matrix $A$, we can deform this matrix integral to
\begin{align}
    Z_{\mathrm{2det}}=\frac{\det_{N}(NY)^{N}}{(2\pi g)^{N^{2}}}\int dK dM_{N\times N}\,e^{-\frac{N}{g}\Tr(\sqrt{Y}K\sqrt{Y}M)-\sum_{k\geq 1}\frac{t_{k}}{k}\Tr(M^{k})-\sum_{k\geq 1}\frac{\bar{t}_{k}}{k}\Tr(K^{k})}\,,
\end{align}
where we introduced \textit{Miwa variables} by
\begin{align}
    s_{k}=\Tr_{N}(Y^{-k})\,,\quad t_{k}=\Tr_{Q}(X^{-k}) \,,\quad \bar{t}_{k}=\Tr_{R}(V^{-k})\,.
\end{align}

To obtain the \textit{F(ace)-Type} dual, we first integrate in the fermionic degrees of freedom by rewriting the determinants. 
Since we can rewrite $\prod_{a=1}^{Q}\det(x_{a}-M) = \det(X \otimes I_N - I_Q \otimes M)$ \cite{Maldacena:2004sn},
we introduce $N\times Q$ complex fermions $\psi^{\dagger}_{ia}$ and $N\times R$ fermions $\chi_{i\mu}^{\dagger}$.
Physically, the fermions $\psi^{\dagger}_{ia}$ represent bifundamental open-string modes stretching between $N$ D3 branes and $Q$ D3' branes.
Similarly, the fermions $\chi^{\dagger}_{i\m}$ represent $N$ D3 branes and $R$ D3'' branes. We then have
\begin{equation}
    \begin{split}
        Z_{\mathrm{2det}}
        \, = \, \frac{\mathcal{K}}{Z_{N}}\int &dK dM d\psi d\psi^{\dagger} d\chi d\chi^{\dagger}
        \exp\bigg[-\frac{N}{g}\Tr_{N}(\sqrt{Y}K\sqrt{Y}M)\\
        &\qquad+\psi_{ia}^{\dagger}(X_{ab}\delta_{ij}-\delta_{ab}M_{ij})\psi_{jb}+\chi_{i\mu}^{\dagger}(V_{\mu\nu}\delta_{ij}-K_{ij}\delta_{\mu\nu})\chi_{j\nu}\bigg]\,,
    \end{split}
\end{equation}
where we defined
\begin{align}
    \mathcal{K}=\left[\frac{\det_{N}(Y)}{\det_{Q}(X)\det_{R}(V)}\right]^{N}\,,\qquad Z_{N}=\left(\frac{2\pi g}{N}\right)^{N^{2}}\,.
\end{align}
We can integrate out the original matrices $K$ and $M$. Since the integration contour for $K$ is on the imaginary axis, the integration over $K$ yields a delta function. We then carry out the integral over $M$ to obtain
\begin{equation}
    \begin{split}
        Z_{\mathrm{2det}}&=\left(\frac{1}{\det_{Q}(X)\det_{R}(V)}\right)^{N}\int d\psi d\psi^{\dagger} d\chi d\chi^{\dagger}\\
        &\qquad\times\exp\left(\psi^{\dagger}_{ia}X_{ab}\psi_{ib}+\chi^{\dagger}_{i\mu}V_{\mu\nu}\chi_{i\nu}-\frac{g}{N}\psi^{\dagger}_{ia}(Y^{-\frac{1}{2}})_{ik}\chi_{k\mu}\chi_{l\mu}^{\dagger}(Y^{-\frac{1}{2}})_{lj}\psi_{ja}\right)\,.
    \end{split}
\end{equation}
We now introduce a complex bosonic $Q\times R$ matrix $S_{a\mu}$. By using the Hubbard–Stratonovich transformation, we obtain
\begin{equation}
    \begin{split}
        Z_{\mathrm{2det}}&=\frac{1}{Z_{QR}(\det_{Q}(X)\det_{R}(V))^{N}}\int d\psi d\psi^{\dagger} d\chi d\chi^{\dagger} dS dS^{\dagger}\\
        &\times\exp\left(\frac{N}{g}S^{\dagger}_{\mu a}S_{a\mu}-S^{\dagger}_{\mu a}\chi^{\dagger}_{l\mu}(Y^{-\frac{1}{2}})_{lj}\psi_{ja}-\psi_{ia}^{\dagger}(Y^{-\frac{1}{2}})_{ik}\chi_{k\mu}S_{a\mu}+\psi^{\dagger}_{ia}X_{ab}\psi_{ib}+\chi^{\dagger}_{i\mu}V_{\mu\nu}\chi_{i\nu}\right)\,, \label{eq:GKMS-HS}
    \end{split}
\end{equation}
with $Z_{QR}=(\pi g/N)^{QR}$. 
We note that this procedure is formal because this integral does not converge. 
Alternatively, we can also use
\begin{align}
    \exp\left(-\frac{N}{g}S^{\dagger}_{\mu a}S_{a\mu}+iS^{\dagger}_{\mu a}\chi^{\dagger}_{l\mu}(Y^{-\frac{1}{2}})_{lj}\psi_{ja}+i\psi_{ia}^{\dagger}(Y^{-\frac{1}{2}})_{ik}\chi_{k\mu}S_{a\mu}+\psi^{\dagger}_{ia}X_{ab}\psi_{ib}+\chi^{\dagger}_{i\mu}V_{\mu\nu}\chi_{i\nu}\right) \label{eq:KS-HS}
\end{align}
as an integrand.
We adopt the convention~\eqref{eq:GKMS-HS} in the two-point function case to align the notation with~\cite{Gopakumar:2024jfq}, but in the three-point case, we use the convention~\eqref{eq:KS-HS}.
Finally, we integrate out the fermionic fields  with the field redefinition
\begin{align}
    S\rightarrow X^{-\frac{1}{2}}SV^{-\frac{1}{2}}\,, \qquad S^{\dagger}\rightarrow V^{-\frac{1}{2}}S^{\dagger}X^{-\frac{1}{2}}
\end{align}
to obtain the F-type open string theory
\begin{equation}
    \begin{split}
        &Z_{\mathrm{2det}}[t_{k},\bar{t}_{k},s_{k}] \, = \, Z_{2F}\\
        &:=\frac{(\det_{Q}(X))^{R}(\det_{R}(V))^{Q}}{\det_{N}(Y)^{Q}Z_{QR}}\int dS dS^{\dagger}_{Q\times R} \, \prod_{i=1}^{N}\det(y_{i}-SS^{\dagger})
        \exp\left[\frac{N}{g}\Tr(VS^{\dagger}XS) \right]\\
        &=\frac{(\det_{Q}(X))^{R}(\det_{R}(V))^{Q}}{Z_{QR}}\int dS dS^{\dagger}_{Q\times R}\exp\left[\frac{N}{g}\Tr(VS^{\dagger}XS)-\sum_{k\geq 1}\frac{s_{k}}{k}\Tr(SS^{\dagger})^{k})\right]\,.\label{eq:2pt-Ftype}
    \end{split}
\end{equation}
Physically, this theory can be understood as the effective theory on the D3'/D3'' open string. 
The above trick was originally introduced for the one-point functions~\cite{Maldacena:2004sn, Brezin:2016eax}.

\subsection{Three-point functions}
\label{sec:threept-MatrixDuality}
Let us next consider the case for three-point functions.
Similar to section~\ref{sec:twopt-MatrixDuality}, we consider an action
\begin{equation}
    \begin{split}
        S&=-\frac{1}{2}\sum_{\alpha=1}^{3}\frac{1}{\kappa_{\alpha}^{2}}\Tr(\sqrt{Y}M_{\alpha}\sqrt{Y}M_{\alpha})+\frac{1}{2}\sum_{\substack{\alpha,\beta=1\\\alpha\neq\beta}}^{3}\frac{1}{\kappa_{\alpha}\kappa_{\beta}}\Tr(\sqrt{Y}M_{\alpha}\sqrt{Y}M_{\beta})\\
        &=\frac{1}{2}\Tr \left(\sqrt{Y} M_{\alpha}K_{\alpha\beta}\sqrt{Y}M_{\beta}\right)\,,
    \end{split}
\end{equation}
where the repeated indices in the second line are implicitly summed, and we introduced the source matrix
\begin{align}
    Y=\mathrm{diag}(y_{1},\ldots,y_{N})\,,
\end{align}
and the kernel 
\begin{align}
    K=\begin{pmatrix}
            -\frac{1}{\kappa_{1}^{2}} & \frac{1}{\kappa_{1}\kappa_{2}} & \frac{1}{\kappa_{1}\kappa_{3}} \\
            \frac{1}{\kappa_{1}\kappa_{2}} & -\frac{1}{\kappa_{2}^{2}} & \frac{1}{\kappa_{2}\kappa_{3}} \\
            \frac{1}{\kappa_{1}\kappa_{3}} & \frac{1}{\kappa_{2}\kappa_{3}} & -\frac{1}{\kappa_{3}^{2}}
        \end{pmatrix}\,.
\end{align}
The partition function can be evaluated formally:
\begin{align}
    Z_{\mathrm{3pt}}=\int dM_{1} dM_{2}dM_{3}\,e^{-\frac{N}{g}S}=\left(\frac{1}{\det_{N} (NY)}\right)^{\frac{3}{2}N}\frac{(2\pi g)^{\frac{3N^{2}}{2}}}{(\det K)^{\frac{N^{2}}{2}}}\,.
\end{align}
We then consider the three-point V-type model defined by the determinant insertion as follows:
\begin{align}
        &Z_{\mathrm{3det}}[t_{1k},t_{2k},t_{3k},s_{k}] \, = \, Z_{3V} \label{eq:V-type}\\
        &:=\mathcal{K}\int dM_{1}dM_{2}dM_{3\,N\times N}
        \, e^{-\frac{N}{g}S}\prod_{a=1}^{Q}\det(u_{a}-M_{1})\prod_{a'=1}^{R}\det(v_{a'}-M_{2})\prod_{\hat{a}=1}^{S}\det(w_{\hat{a}}-M_{3})\nn\\
        &=\tilde{\mathcal{K}}\int dM_{1}dM_{2}dM_{3\,N\times N}\,e^{-\frac{N}{g}S-\sum_{k\geq 1}\frac{t_{1k}}{k}\Tr M_{1}^{k}-\sum_{k\geq 1}\frac{t_{2k}}{k}\Tr M_{2}^{k}-\sum_{k\geq 1}\frac{t_{3k}}{k}\Tr M_{3}^{k}}\,, \nn 
\end{align}
where
\begin{equation}
    \mathcal{K}=\left(\frac{\det(NY)^{\frac{3}{2}}}{\det_{Q} U \det_{R} V \det_{S} W}\right)^{N}\frac{(\det K)^{\frac{N^{2}}{2}}}{(2\pi g)^{\frac{3N^{2}}{2}}}\,,
\end{equation}
and
\begin{equation}
    \widetilde{\mathcal{K}}=\mathcal{K}\,(\mathrm{det}_{Q}U\,\mathrm{det}_{R}V\,\mathrm{det}_{S}W)^{N}
\end{equation}
with diagonal source matrices
\begin{equation}
    U=\mathrm{diag}(u_{1},\ldots,u_{Q})\,,\quad V=\mathrm{diag}(v_{1},\ldots,v_{R})\,,\quad W=\mathrm{diag}(w_{1},\ldots,w_{S})\,,
\end{equation}
and Miwa variables
\begin{align}
    t_{1k}=\Tr\,U^{-k}\,,\qquad t_{2k}=\Tr\,V^{-k}\,,\qquad t_{3k}=\Tr\,W^{-k}\,,\qquad
    s_{k}=\Tr\,Y^{-k}\,.
\end{align}

Let us deduce the F-type string dual in this three-point case. Our strategy is very much similar to the two-point case.
We first rewrite the determinants in terms of the fermion path integrals to obtain
\begin{align}
        Z_{\mathrm{3det}}&=\mathcal{K}\int dM_{1}dM_{2}dM_{3}d\eta d\eta^{\dagger} d\psi d\psi^{\dagger} d\chi d\chi^{\dagger}\exp\bigg[-\frac{N}{g}S+\eta^{\dagger}_{ia}(U_{ab}\delta_{ij}-\delta_{ab}M_{1,ij})\eta_{jb} \nn\\
        &\qquad+\psi^{\dagger}_{ia'}(V_{a'b'}\delta_{ij}-\delta_{a'b'}M_{2,ij})\psi_{jb'}+\chi^{\dagger}_{i\hat{a}}(W_{\hat{a}\hat{b}}\delta_{ij}-\delta_{\hat{a}\hat{b}}M_{3,ij})\chi_{j\hat{b}}\bigg] \, . \label{eq:integrating-in-fermion}
\end{align}
By completing the square, we can then carry out the Gaussian integral for the Hermitian matrices $M_i$:
\begin{align}
    \begin{split}
        Z_{\mathrm{3det}}
        &=\frac{1}{(\det_{Q}U)^{N}(\det_{R}V)^{N}(\det_{S}W)^{N}}\int d\eta d\eta^{\dagger} d\psi d\psi^{\dagger} d\chi d\chi^{\dagger} \\
        &\qquad \times \exp\Big(\eta^{\dagger}_{ia}U_{ab}\eta_{ib}+\psi^{\dagger}_{ia'}V_{a'b'}\psi_{ib'}
        + \chi^{\dagger}_{i\hat{a}}W_{\hat{a}\hat{b}}\chi_{i\hat{b}}+\frac{g}{2N}\Tr (B_{\alpha}K^{-1}_{\alpha\beta}B_{\beta})\Big)
    \end{split}
\end{align}
with
\begin{align}
    K^{-1}=\frac{1}{2}\begin{pmatrix}
        0 & \kappa_{1}\kappa_{2} & \kappa_{1}\kappa_{3}\\
        \kappa_{2}\kappa_{1} & 0 & \kappa_{2}\kappa_{3}\\
        \kappa_{3}\kappa_{1}& \kappa_{3}\kappa_{2} & 0
      \end{pmatrix}
\end{align}
and
\begin{align}
    (B_{1})_{ij}=Y^{-\frac{1}{4}}_{ik}\eta_{ka}\eta^{\dagger}_{la}Y^{-\frac{1}{4}}_{lj}\,,\qquad (B_{2})_{ij}=Y^{-\frac{1}{4}}_{ik}\psi_{ka'}\psi^{\dagger}_{la'}Y^{-\frac{1}{4}}_{lj}\,,\qquad (B_{3})_{ij}=Y^{-\frac{1}{4}}_{ik}\chi_{k\hat{a}}\chi^{\dagger}_{l\hat{a}}Y^{-\frac{1}{4}}_{lj}\,.
\end{align}
For the detailed manipulations, see Appendix~\ref{sec:completing-square}.

Let us next integrate in rectangular bifundamental bosonic matrices $\Sigma_{i}$ and $\Sigma_{i}^{\dagger}$. As we will see, the matrix $\S_{1}$ is $Q\times R$, $\S_{2}$ is $R\times S$, and $\S_{3}$ is $S\times Q$ matrix. We introduce the Gaussian integrals to obtain
\begin{align}
    Z_{\mathrm{3det}}=\mathcal{K}_{3}\int \prod_{i=1}^{3} d\Sigma_{i}d\Sigma_{i}^{\dagger}\,d\eta d\eta^{\dagger} d\psi d\psi^{\dagger} d\chi d\chi^{\dagger}\exp\left(\widetilde{I}[\Sigma_{i},\Sigma_{i}^{\dagger},\eta,\psi,\chi]\right)\,,\label{eq:interating-in-boson}
\end{align}
where
\begin{align}
    \mathcal{K}_{3}=\frac{1}{(\det_{Q}U)^{N}(\det_{R}V)^{N}(\det_{S}W)^{N}}\left(\frac{N}{2\pi g}\right)^{QR+RS+SQ}\,,
\end{align}
\begin{equation}
    \begin{split}
    \widetilde{I}[\Sigma_{i},\Sigma_{i}^{\dagger},\eta,\psi,\chi]&=-\frac{N}{2g}\left[\Tr(\Sigma_{1\,a'a}^{\dagger}\Sigma_{1\,aa'}+\Sigma_{2\,\hat{a}a'}^{\dagger}\Sigma_{2\,a'\hat{a}}+\Sigma_{3\,a\hat{a}}^{\dagger}\Sigma_{3\,\hat{a}a})\right]\\
    &\qquad+\sum_{l}
    \begin{pmatrix}
        \eta_{la}^{\dagger} & \psi_{la'}^{\dagger} & \chi_{l\hat{a}}^{\dagger} 
    \end{pmatrix}
    \,A_{l\{a,a',\hat{a}\}\{b,b',\hat{b}\}}\,
    \begin{pmatrix}
        \eta_{lb} \\ \psi_{lb'} \\ \chi_{l\hat{b}} 
    \end{pmatrix}\,,
    \end{split}
\end{equation}
and 
\begin{align}
    A_{l}=
    \begin{pmatrix}
        U_{ab} & \frac{i}{2}\sqrt{\kappa_{1}\kappa_{2}}y^{-\frac{1}{2}}_{l}\Sigma_{1\,ab'} & \frac{i}{2}\sqrt{\kappa_{1}\kappa_{3}}y^{-\frac{1}{2}}_{l}\Sigma^{\dagger}_{3\,a\hat{b}} \\
        \frac{i}{2}\sqrt{\kappa_{1}\kappa_{2}}y^{-\frac{1}{2}}_{l}\Sigma^{\dagger}_{1\,a'b} & V_{a'b'} &  \frac{i}{2}\sqrt{\kappa_{2}\kappa_{3}}y^{-\frac{1}{2}}_{l}\Sigma_{2\,a'\hat{b}}\\
        \frac{i}{2}\sqrt{\kappa_{1}\kappa_{3}}y^{-\frac{1}{2}}_{l}\Sigma_{3\,\hat{a}b} & \frac{i}{2}\sqrt{\kappa_{2}\kappa_{3}}y^{-\frac{1}{2}}_{l}\Sigma^{\dagger}_{2\,\hat{a}b'}& W_{\hat{a}\hat{b}}
    \end{pmatrix}\,.
\end{align}
For the detailed calculations, see Appendix~\ref{sec:integrate-in}.
Since the matrix $A_{l}$ factorizes into the product of a block diagonal matrix and the rest, we have
\begin{align}
    A_{l}=\mathrm{diag}(U,V,W)\left(I_{(Q+R+S)\times(Q+R+S)}+\frac{i}{2}\,y_{l}^{-\frac{1}{2}}\,\Lambda\right)\,,
\end{align}
where
\begin{align}
   \Lambda=
    \begin{pmatrix}
        0 & \sqrt{\kappa_{1}\kappa_{2}}U^{-1}\Sigma_{1} & \sqrt{\kappa_{1}\kappa_{3}}U^{-1}\Sigma^{\dagger}_{3} \\
        \sqrt{\kappa_{1}\kappa_{2}}V^{-1}\Sigma^{\dagger}_{1} & 0 & \sqrt{\kappa_{2}\kappa_{3}}V^{-1}\Sigma_{2}\\
        \sqrt{\kappa_{1}\kappa_{3}}W^{-1}\Sigma_{3} & \sqrt{\kappa_{2}\kappa_{3}}W^{-1}\Sigma_{2}^{\dagger} & 0
    \end{pmatrix}\,. 
\end{align}
After integrating out the fermions, we then obtain the F-type open string description
\begin{equation}
    \begin{split}
        Z_{\mathrm{3det}} \, = \, Z_{3F}
        \, &:= \, \left(\frac{N}{2\pi g }\right)^{QR+RS+SQ}\int \prod_{i=1}^{3}\, d\Sigma_{i}d\Sigma_{i}^{\dagger}\exp\left[-\sum_{k=1}^{\infty}\frac{1}{k}\left(-\frac{i}{2}\right)^{k}s_{\frac{k}{2}}\Tr\Lambda^{k}\right]\\
        &\qquad\qquad\times\exp\left\{-\frac{N}{2g}\Big[\Tr(\Sigma_{1}^{\dagger}\Sigma_{1}+\Sigma_{2}^{\dagger}\Sigma_{2}+\Sigma_{3}^{\dagger}\Sigma_{3})\Big]\right\}\,.
    \end{split} 
\end{equation}
Here, we extend the definition of the Miwa variable to half-integers:
\begin{align}
    s_{\frac{k}{2}}=\Tr Y^{-\frac{k}{2}}\,.
\end{align}

We can also rewrite this F-type model as closed quiver walks.
The trace of $\Lambda^{n}$ can be simplified by
\begin{equation}
    \Tr\Lambda^{n}=\sum_{\substack{i_1,\ldots,i_n\in\{1,2,3\}\\ i_{r+1}\ne i_r,\; i_{n+1}=i_1}}\Tr(X_{i_{1} i_{2}}X_{i_{2} i_{3}}\ldots X_{i_{n} i_{1}})\,,
\end{equation}
where
\begin{align}
    &X_{12}=\sqrt{\kappa_{1}\kappa_{2}}U^{-1}\Sigma_{1}\,,\qquad X_{13}=\sqrt{\kappa_{1}\kappa_{3}}U^{-1}\Sigma_{3}^{\dagger}\,,\nn\\
        &X_{21}=\sqrt{\kappa_{1}\kappa_{2}}V^{-1}\Sigma_{1}^{\dagger}\,,\qquad X_{23}=\sqrt{\kappa_{2}\kappa_{3}}V^{-1}\Sigma_{2}\,,\\
        &X_{31}=\sqrt{\kappa_{1}\kappa_{3}}W^{-1}\Sigma_{3}\,,\qquad X_{32}=\sqrt{\kappa_{2}\kappa_{3}}W^{-1}\Sigma_{2}^{\dagger}\,.\nn
\end{align}
Then, the F-type model is given by
\begin{equation}
    \begin{split}
        Z_{3F} \, &=  \left(\frac{N}{2\pi g}\right)^{QR+RS+SQ}\int \prod_{i=1}^{3}\, d\Sigma_{i}d\Sigma_{i}^{\dagger} \, \exp\left[-\frac{N}{2g}\Tr\Big(\Sigma_{1}^{\dagger}\Sigma_{1}+\Sigma_{2}^{\dagger}\Sigma_{2}+\Sigma_{3}^{\dagger}\Sigma_{3}\Big)\right] \\
        &\qquad \times\exp\left[-\sum_{k=1}^{\infty}\frac{1}{k}\left(-\frac{i}{2}\right)^{k}s_{\frac{k}{2}}\sum_{\substack{i_1,\ldots,i_k\in\{1,2,3\}\\ i_{r+1}\ne i_r,\; i_{k+1}=i_1}}\Tr(X_{i_{1} i_{2}}X_{i_{2} i_{3}}\ldots X_{i_{k} i_{1}})\right] \, .
    \end{split}
\end{equation}

\section{Single Giant Correlators}
\label{sec:single}
The large $N$ limit of the $\mathcal{N}=4$ SYM single giant graviton correlators was discussed in \cite{Jiang:2019xdz}.
In this section, we show that our F-type models indeed agree with their model in the single giant graviton cases ($Q=R=S=1$).
On the other hand, we will show that our V-type models are exactly solvable for arbitrary $N$, when we restrict to the single giant graviton cases.

\subsection{F-type two-point functions}
\label{sec:f2}
We start from the F-type model of the two-point function~\eqref{eq:2pt-Ftype}. The single determinant correlator corresponds to 
	\begin{align}
		Q \, = \, 1, \qquad R \, = \, 1 \, ,\qquad X_{Q\times Q} \, = \, x \, , \qquad V_{R \times R} \, = \, v \, ,
	\end{align}
and the F-type model is reduced to
\begin{align}
		Z_{\mathrm{2F}} \, = \, \frac{xv}{\det_N Y} \left( \frac{N}{\pi g} \right) \int ds ds^* \exp\left[ \frac{xvN}{g} |s|^2 \, + \, \sum_{i=1}^N \, \log \Big( y_i - |s|^2 \Big) \right] \, ,
\end{align}
where $|s|^{2}=ss^{*}$.
In order to make the integrals convergent, we can rotate the integration contour by $s=i \s$ and $s^* = i \s^*$. Then,  we take
	\begin{align}
		\s \, = \, (a+ib) \,, \qquad \s^{*} \, = \, a-ib
	\label{eq:countor-rot}
	\end{align}
with real variables $a,b$ for the convergence of the integrals as in~\cite{Jiang:2019xdz}.
When the source is uniform $y_i =y$, we can further simplify to
\begin{align} 
		Z_{\mathrm{2F}}  \, = \, \frac{Nxv}{\pi g}  \int d\s d\s^* \exp\big[ N S_{\eff} \Big] \, ,
	\end{align}
where 
	\begin{align}
		S_{\eff} \, = \, - \frac{xv}{g} \, |\s|^2 \, + \, \log\left(1+ \frac{|\s|^2}{y}\right) \, .
	\end{align}
In fact, now the integrals can be evaluated exactly.
By defining 
	\begin{align}
		q_y \, := \, \frac{g}{xvy} \, , \qquad z \, := \, \frac{|\s|^2}{y} \, , 
	\end{align}
we can rewrite 
	\begin{align} 
		Z_{\mathrm{2F}} \, &= \, \frac{N}{q_y} \int_0^\inf dz \, \exp\left[ N \left( \log(1+z) \, - \, \frac{z}{q_y} \right) \right] \nn\\
		&= \, \frac{N}{q_y} \sum_{r=0}^N \binom{N}{r} \int_0^\inf dz \, z^r \, e^{-Nz/q_y} \, .
	\end{align}
Now we evaluate the integral term by term. This leads to
	\begin{align} 
		Z_{\mathrm{2F}} \, &= \, \frac{N}{q_y} \sum_{r=0}^N \binom{N}{r} r! \left( \frac{q_y}{N} \right)^{r+1}
		\, = \, \sum_{r=0}^N \frac{(N)_r}{N^r} q_y^r \, ,
	\end{align}
where $(N)_r=\G(N+1)/\G(N-r+1)$.

We can also evaluate the large $N$ limit.  The large $N$ saddle-point solution $\s=\s_{0}$ is given by 
\begin{align} 
		|\s_0|^2 \, = \, \frac{g}{xv} \, - \, y \, ,
\end{align}
and the tree-level partition function is 
\begin{align}
    Z_{\tree} \, = \, \frac{q_y^{N-1} N}{\pi y} \, e^{N(\frac{xvy}{g} - 1)} \, .
	\label{eq:Z_tree-2F}
\end{align}
Here, we assumed the condition
\begin{align}
    q_{y}>1
\end{align}
for the existence of the saddle.
The one-loop contributions can also be evaluated. For this purpose, we change the integration variables by 
	\begin{align}
		\s \, = \, r e^{i \th} \, .
	\end{align}
and then, 
the saddle-point solution $r=r_{0}$ with any $\theta$ corresponds to 
    \begin{align}
		r_0^2 \, = \, \frac{g}{xv} \, - \, y \, .
	\end{align}
We consider the fluctuation around the saddle:
\begin{align}
    r=r_{0}+\delta r\,.
\end{align}
The one-loop contribution to the partition function is now written as
    \begin{align}
		Z_{\oneloop} \, &= \, 2\pi r_0 \int_{-\infty}^\infty d\d r \exp \left[ -\frac{2xvN}{g}\left( 1 - \frac{xvy}{g} \right) \d r^2 \right] \nn\\[4pt]
		&= \, \frac{\pi^{\frac{3}{2}}g}{xv}\sqrt{\frac{2}{N}} \, .
	\label{eq:Z_oneloop-2F}
	\end{align}

In \cite{Jiang:2019xdz}, sourceless single determinant correlators were studied.
In order to study the sourceless limit ($x \to 0$, $v \to 0$), it is convenient to rescale 
	\begin{align}
		s \, \to \, \frac{\r}{x} \, ,\qquad s^* \, \to \, \frac{\r^*}{v} \, ,
	\end{align}
and then send $x \to 0$, $v \to 0$. This leads to the effective action
	\begin{align}
		S_{\eff} \, = \, \frac{|\r|^2}{g} \, + \, \log (-|\r|^2) \, ,
	\end{align}
where we simply dropped  the term since we can absorb the term involving sources in the overall normalization. 
This agrees with the effective action for the two-point function studied in \cite{Jiang:2019xdz}.\footnote{When we denote the coupling $g^{2}$ in~\cite{Jiang:2019xdz} by $g^{2}_{\mathrm{JKV}}$, our coupling is related to $g^{2}_{\mathrm{JKV}}=\frac{gw_{12}^{2}}{4Y_{1}\cdot Y_{2}}$. We further rescale $\rho$ to obtain the result in~\cite{Jiang:2019xdz}.}

\subsection{V-type two-point functions}
\label{sec:v2}
Let us define the V-type two-point function by
	\begin{align}
		Z_{Q,R}^{(N)}(X,V) \, = \, Z_{2V} \, = \, \frac{\left\la \prod_{a=1}^Q \det_N(x_a - M) \prod_{a'=1}^R \det_N(v_{a'} - K) \right\ra_{K,M}}{\det_Q(X)^N \det_R(V)^N} \, ,
	\end{align}
where 
	\begin{align}
		\big\la \mathcal{O} \big\ra_{K,M} \, = \, \frac{y^{N^2}}{Z_{N}}\color{black} \int dK \, dM \, e^{-\frac{yN}{g} \Tr_N(KM)} \, \mathcal{O} \, .
	\end{align}
For simplicity, we took the uniform source $Y=y \cdot \mathds{1}$.
We note that the propagator is given by
	\begin{align}
		\big\la K_{ij} M_{kl} \big\ra_{K,M} \, = \, \frac{g}{yN} \, \d_{il} \d_{jk} \, .\label{eq:KM-propagator}
	\end{align}
	
Now we consider the single giant case $Q=R=1$. This is written as
	\begin{align}
		Z_{1,1}^{(N)}(x,v) \, = \,  \left\la \det\left(\mathds{1} - \frac{M}{x}\right) \det\left(\mathds{1} - \frac{K}{v}\right) \right\ra_{K,M} \, .
	\end{align}
Hereafter, we will omit the subscript $N$ for the determinants.
We can find an exact expression for this $Z_{1,1}^{(N)}$ for a finite $N$.
First, we expand both determinants into elementary symmetric polynomials:
	\begin{align}
		\det\!\left(\mathds{1}-\frac{M}{x}\right) \, = \, \sum_{r=0}^{N} (-1)^r x^{-r} e_r(M) \, , \qquad
		\det\!\left(\mathds{1}-\frac{K}{v}\right) \, = \, \sum_{s=0}^{N} (-1)^s v^{-s} e_s(K) \, ,
	\label{eq:elementary-symmetric-polynomials}
	\end{align}
with
	\begin{align}
		e_r(M) \, = \, \sum_{\substack{I\subset\{1,\ldots,N\}\\ |I|=r}} \det M_I,
	\end{align}
where $M_I$ is the $r\times r$ principal submatrix of $M$ with rows and columns restricted to the same subset $I$.
Substituting into the two-point function gives
	\begin{align}
		Z_{1,1}^{(N)}(x,v) \, = \, \sum_{r,s=0}^{N} (-1)^{r+s} x^{-r} v^{-s} \big\langle e_r(M)e_s(K) \big\rangle_{K,M} \, .
	\end{align}
Since the only nonzero propagator connects one $M$ to one $K$ as in~\eqref{eq:KM-propagator}, the correlator  vanishes
	\begin{align}
		\left\langle e_r(M)e_s(K)\right\rangle \, = \, 0    
	\end{align}
    when $r\neq s$.
Therefore, we have
	\begin{align}
		Z_{1,1}^{(N)}(x,v) \, = \, \sum_{r=0}^{N} \, (xv)^{-r} \big\langle e_r(M)e_r(K) \big\rangle_{K,M} \, .
	\end{align}
Now we expand the elementary symmetric polynomials into principal minors.
Because of the propagator, only diagonal pairs of principal minors contribute:
	\begin{align}
		\big\langle e_r(M)e_r(K) \big\rangle_{K,M} \, = \,  \sum_{\substack{I\subset\{1,\ldots,N\}\\ |I|=r}} \big\langle \det M_I\,\det K_I \big\rangle_{K,M} \, .
	\end{align}
For a fixed $r$-element subset $I$, let us define
	\begin{align}
		A \, = \, M_I \, , \qquad B \, = \, K_I \, .
	\end{align}
Then, expanding both determinants,
	\begin{align}
		\det A \, = \, \sum_{\sigma\in S_r} \sgn(\sigma) \prod_{\alpha=1}^{r} A_{\alpha,\sigma(\alpha)} \, , \qquad
		\det B \, = \, \sum_{\tau\in S_r} \sgn(\tau) \prod_{\alpha=1}^{r} B_{\alpha,\tau(\alpha)} \, .
	\end{align}
Since, for a fixed $\sigma$,  the nonzero contribution from $\det B$ has $\tau=\sigma^{-1}$ from the propagator~\eqref{eq:KM-propagator},  we have
	\begin{align}
		\sgn(\sigma)\sgn(\sigma^{-1}) \, = \, 1.
	\end{align}
There are $r!$ such choices, each giving $r$ propagators, and there are $\binom{N}{r}$ principal minors. Hence, we have
	\begin{align}
		\left\langle e_r(M)e_r(K)\right\rangle \, = \,  \binom{N}{r} r!\left(\frac{g}{yN}\right)^r \, = \, (N)_r\left(\frac{g}{yN}\right)^r \, .
	\end{align}
Therefore, we find 
	\begin{align}
		Z_{1,1}^{(N)}(x,v) \, = \, \sum_{r=0}^{N} \frac{(N)_r}{N^r} \, q_y^r \, ,
	\label{eq:Z_{1,1}}
	\end{align}
where
	\begin{align}
		q_y \, = \, \frac{g}{xvy} \, .
	\end{align}

Let us also consider the large $N$ limit of this single giant graviton correlator.
The result from the F-type model up to one-loop~\eqref{eq:Z_tree-2F} and~\eqref{eq:Z_oneloop-2F} is schematically $Z \sim \sqrt{2\pi N} q_y^N e^{-N}$ for large $q_y$.
This result corresponds to the top $r=N$ contribution in the summation in (\ref{eq:Z_{1,1}}), which gives
	\begin{align}
		Z_{1,1}^{(N)}(x,v) \, \sim \, \frac{(N)_N}{N^N} \, q_y^N \, = \, \frac{N!}{N^N} \, q_y^N \, \approx \, \sqrt{2\pi N} \, q_y^N e^{-N} \, ,
	\end{align}
where in the last step we used Stirling's approximation. This result agrees with the large $N$ expansion of the F-type.

\subsection{F-type three-point functions}
\label{sec:f3}
Now we study the single-determinant three-point functions, which correspond to
	\begin{align}
		Q \, = \, 1, \qquad R \, = \, 1 \, , \qquad S \, = \, 1 \, , \qquad U_{Q\times Q} \, = \, u \, , \qquad V_{R \times R} \, = \, v \, , \qquad W_{S \times S} \, = \, w \, ,
	\end{align}
and 
\begin{equation}
    \begin{split}
        Z_{\mathrm{3F}} \, 
        &= \, \left(\frac{N}{2\pi g}\right)^{3} \int \prod_{\a=1}^3 d\s_\a d\s_\a^* \exp\left[ - \frac{N}{2g} \bigg( |\s_1|^2+ |\s_2|^2+ |\s_3|^2 \right)
		\,\\
        &\hspace{6 cm}+ \, \sum_{i=1}^N \, \Tr_3 \log \Big( \mathds{1} + \frac{iy_i^{-\frac{1}{2}}}{2} \L \Big) \bigg] \, ,
    \end{split}
\end{equation}
where $\mathds{1}$ is the identity matrix, and
	\begin{align}
		\L \, = \, 
		\begin{pmatrix}
			0 & \frac{\sqrt{\kappa_{1}\kappa_{2}}\s_1}{u} & \frac{\sqrt{\kappa_{1}\kappa_{3}}\s_3^*}{u} \\
			\frac{\sqrt{\kappa_{1}\kappa_{2}}\s_1^*}{v} & 0 & \frac{\sqrt{\kappa_{2}\kappa_{3}}\s_2}{v} \\
			\frac{\sqrt{\kappa_{1}\kappa_{3}}\s_3}{w} & \frac{\sqrt{\kappa_{2}\kappa_{3}}\s_2^*}{w} & 0 
		\end{pmatrix} \, .
	\end{align}
When the source is uniform $y_i =1$, we can further simplify to\footnote{The result with the uniform source $y_i=y$ can be recovered by rescaling the coupling $g \to g/y$ as in the two-point functions.}
	\begin{align}
		Z_{\mathrm{3F}} \, = \,\, \left(\frac{N}{2\pi g}\right)^{3}\int \prod_{\a=1}^3 d\s_\a d\s_\a^* \exp\big[ N S_{\eff} \Big] \, ,
	\end{align}
where 
	\begin{align}
		S_{\eff} \, = \, - \frac{1}{2g} \left( |\s_1|^2+ |\s_2|^2+ |\s_3|^2 \right)
		+ \, \Tr_3 \log \Big( \mathds{1} + \frac{i}{2} \L \Big) \, .
	\end{align}
In this case, we can use the large $N$ limit. The large $N$ saddle-point equations are given by 
	\begin{align}
		\frac{\s_1}{2g} \, &= \, - \, \frac{1}{\det(\mathds{1} + \frac{i}{2} \L)} \left( \frac{i\k_{1}\k_{2}\k_{3}}{8uvw}\s_2^* \s_3^* \, - \, \frac{\k_{1}\k_{2}}{4uv}\s_1 \right) \, , \\
		\frac{\s_2}{2g} \, &= \, - \, \frac{1}{\det(\mathds{1} + \frac{i}{2} \L)} \left( \frac{i\k_{1}\k_{2}\k_{3}}{8uvw}\s_3^* \s_1^* \, - \, \frac{\k_{2}\k_{3}}{4vw}\s_2 \right) \, , \\
		\frac{\s_3}{2g} \, &= \, - \, \frac{1}{\det(\mathds{1} + \frac{i}{2} \L)} \left( \frac{i\k_{1}\k_{2}\k_{3}}{8uvw}\s_1^* \s_2^* \, - \, \frac{\k_{1}\k_{3}}{4wu}\s_3 \right) \, .
	\end{align}
These equations are rather complicated to solve exactly.

Let us now consider the source-less limit ($u,v,w \to 0$) as discussed in \cite{Jiang:2019xdz}. 
In the current discussion, we are considering the situation after the rotation of the integration path like~\eqref{eq:countor-rot}.
To compare the result with~\cite{Jiang:2019xdz}, we consider the situation before the rotation of the integration path.
Thus, we carry out the  replacement
\begin{align}
    \sigma_{i}\rightarrow\frac{1}{i}\tilde{\sigma}_{i}\,,\quad \sigma_{i}^{*}\rightarrow\frac{1}{i}\tilde{\sigma}_{i}^{*}\,.
\end{align}
The effective action of the sourceless limit thus becomes
	\begin{align}
		S_{\eff} \, = \,  \frac{1}{2g} \left( |\tilde{\s}_1|^2+ |\tilde{\s}_2|^2+ |\tilde{\s}_3|^2  \right) \, + \, \Tr_3 \log \frac{1}{2}\tilde{\L} \, ,
	\end{align}
where 
	\begin{align}
		\tilde{\L} \, = \, 
		\begin{pmatrix}
			0 & \sqrt{\k_{1}\k_{2}}\tilde{\s}_1 & \sqrt{\k_{1}\k_{3}}\tilde{\s}_3^* \\
			\sqrt{\k_{1}\k_{2}}\tilde{\s}_1^* & 0 & \sqrt{\k_{2}\k_{3}}\tilde{\s}_2 \\
			\sqrt{\k_{1}\k_{3}}\tilde{\s}_3 & \sqrt{\k_{2}\k_{3}}\tilde{\s}_2^* & 0 
		\end{pmatrix} \, .
	\end{align}
As before, we simply drop the term $\log\left(\frac{1}{uvw}\right)$.
Then, the large $N$ saddle-point equations are
	\begin{align}
		&\frac{\tilde{\s}_1}{2g} \, =  \,- \frac{\tilde{\s}_2^* \tilde{\s}_3^*}{\tilde{\s}_{1}\tilde{\s}_{2}\tilde{\s}_{3}+\tilde{\s}_{1}^{*}\tilde{\s}_{2}^{*}\tilde{\s}_{3}^{*}} \, , \\
		&\frac{\tilde{\s}_2}{2g} \, =  \,- \frac{\tilde{\s}_3^* \tilde{\s}_1^*}{\tilde{\s}_{1}\tilde{\s}_{2}\tilde{\s}_{3}+\tilde{\s}_{1}^{*}\tilde{\s}_{2}^{*}\tilde{\s}_{3}^{*}} \, , \\
		&\frac{\tilde{\s}_3}{2g} \, =  \,- \frac{\tilde{\s}_1^* \tilde{\s}_2^*}{\tilde{\s}_{1}\tilde{\s}_{2}\tilde{\s}_{3}+\tilde{\s}_{1}^{*}\tilde{\s}_{2}^{*}\tilde{\s}_{3}^{*}} \, .
	\end{align}
These equations agree with those for the three-point function studied in \cite{Jiang:2019xdz}.\footnote{To compare our result with that of~\cite{Jiang:2019xdz}, we need to take appropriate integration path in~\cite{Jiang:2019xdz} as discussed in section~\ref{sec:f2} for the two-point case. We further carry out the field redefinition to match the coupling constant to obtain our result.}
In terms of the original variables, the solutions are given by
	\begin{align}
		|\s_1|^2 \, = \, |\s_2|^2 \, = \, |\s_3|^2 \, = \, g \, , \qquad \s_1 \s_2 \s_3 \, = \, \s_1^* \s_2^* \s_3^* \, = \, \ve g^{3/2} \, ,
	\end{align}
with $\ve = \pm 1$. This means that if we write $\s_i = \sqrt{g} e^{i \th_i}$, we have a constraint 
	\begin{align}
		\th_1 + \th_2 + \th_3 \, = \, 
		\begin{cases}
			\ 0 \hspace{6pt} (\textrm{mod}\ 2\pi) \quad \textrm{for} \ \ \ve = +1 \\
			\ \pi \hspace{5pt} (\textrm{mod}\ 2\pi) \quad \textrm{for} \ \ \ve = -1 \\
		\end{cases}
	\end{align}
Assuming $N$ is an even integer, then the tree-level partition function is obtained as
	\begin{align}
		Z_{\tree} \, = \, 2 (-1)^{N/2} \left( \frac{N}{2\pi g} \right)^3 \left( \frac{\k_1\k_2\k_3 g^{3/2}}{4} \right)^N \, e^{-3N/2} \, .
	\label{eq:Z_tree-3F}
	\end{align}
We can also evaluate the one-loop contribution. We use the polar coordinates $\s_i = r_i e^{i \th_i}$ and define $\Th:=\th_1+\th_2+\th_3$. We expand the fields around the saddle as
\begin{align}
    r_i = \sqrt{g}+\r_i\,,\qquad \Th=\arccos\ve + \vp\,.
\end{align} 
Then, the one-loop contribution is given by
    \begin{align}
		Z_{\oneloop} \, &= \, 4\pi^2 g^{3/2} \int_{\mathds{R}^3} d^3 \r \, e^{-\frac{N}{g} (\r_1^2+\r_2^2+\r_3^2)} \int_{\mathds{R}} d\vp \, e^{-\frac{N}{2} \vp^2} 
		= \, \frac{4\sqrt{2} \pi^4 g^3}{N^2} \, .
	\label{eq:Z_oneloop-3F}
	\end{align}

\subsection{V-type three-point functions}
\label{sec:v3}
Let us define the V-type three-point function by 
	\begin{align}
		Z_{Q,R,S}^{(N)}(U,V,W)
		\, = \, \frac{\left\la \prod_{a=1}^Q \det_N(u_a - M_1) \prod_{a'=1}^R \det_N(v_{a'} - M_2) \prod_{\tilde{a}=1}^S \det_N(w_{\tilde{a}} - M_3) \right\ra_{M^3}}{\det_Q(U)^N \det_R(V)^N \det_S(W)^N} \, ,
	\end{align}
where 
	\begin{align}
		\big\la \mathcal{O} \big\ra_{M^3} \, &=\widetilde{\mathcal{K}} \, \int dM_1 \, dM_2 \, dM_3 \ \mathcal{O}\nn \\
        &\times\ \exp\left[-\frac{N}{2g} \left(- \sum_{\a=1}^3 \frac{1}{\k_{\alpha}^{2}}\Tr_N\big(M_\a^2\big) \, + \, \sum_{\a\ne\b}\frac{1}{\k_{\alpha}\k_{\beta}} \Tr_N\big(M_\a M_\b \big)  \right) \right] \, ,
	\end{align}
We note that the propagator is given by
	\begin{align}
		\big\la (M_\a)_{ij} (M_\b)_{kl} \big\ra_{M^3} \, = \, \frac{G_{\a\b}}{N} \, \d_{il} \d_{jk} \, ,
	\end{align}
with
	\begin{align}
		G_{\a\a} \, = \, 0 \, , \qquad \qquad G_{\a\b} \, = \,  \frac{g}{2}\k_{\alpha}\k_{\beta} \quad (\a \, \ne \b)\,.
	\end{align}

Now, we consider the single giant case $Q=R=S=1$. This is written as
	\begin{align}
		Z_{1,1,1}^{(N)}(u,v,w)
		\, = \, \left\la \det \left(1 - \frac{M_1}{u} \right) \det \left( 1 - \frac{M_2}{v} \right) \det \left( 1 - \frac{M_3}{w} \right) \right\ra_{M^3} \, ,
	\end{align}
where we omitted the subscript $N$ for the determinants.
In the following discussion, it is also convenient to introduce 	
	\begin{align}
		x \, := \, \frac{G_{12}}{uv} \, , \qquad y \, := \, \frac{G_{23}}{vw} \, , \qquad z \, := \, \frac{G_{31}}{wu} \, .
	\end{align}
Using the expansion of the determinants in (\ref{eq:elementary-symmetric-polynomials}), we can rewrite as
	\begin{align}
		Z_{1,1,1}^{(N)}(u,v,w)
		\, = \, \sum_{r_1,r_2,r_3=0}^{N} (-1)^{r_1+r_2+r_3} u^{-r_1} v^{-r_2} w^{-r_3} \big\langle e_{r_1}(M_1) e_{r_2}(M_2) e_{r_3}(M_3) \big\rangle_{M^3} \, .
	\label{eq:Z_{1,1,1}}
	\end{align}
Now we need to evaluate the Wick contractions in $\big\langle e_{r_1}(M_1) e_{r_2}(M_2) e_{r_3}(M_3) \big\rangle_{M^3}$.
Since there is no self-contraction $G_{\a\a}=0$, we introduce $m,n,l$ as the number of contractions between $M_{1}$ and $M_{2}$, $M_{2}$ and $M_{3}$, and $M_{3}$ and $M_{1}$, respectively.
We note that these integers satisfy 
	\begin{align}
		r_1 \, = \, m + l \, \qquad r_2 \, = \, m + n \, , \qquad r_3 \, = \, n + l \, . 
	\end{align}
Since the sum $r_1+r_2+r_3$ is always even, the sign in~\eqref{eq:Z_{1,1,1}} is always +1.
We can further decompose the index contractions. We define $A_{\a\b}$ by a set of index contractions between column index of $M_{\a}$ and row index of $M_{\b}$ for $\a,\b=1,2,3$. For simplicity, we denote
\begin{align}
    A=A_{12}\,,\quad B=A_{21}\,,\quad C=A_{23}\,,\quad D=A_{32}\,,\quad E=A_{31}\,,\quad F=A_{13}\,.
\end{align}
Since each minor is principal, the row-label set and column-label set for each matrix must agree. This gives constraints
	\begin{align}
 		A \cup F \, = \, B \cup E \, , \qquad B \cup C \, = \, A \cup D \, , \qquad D \cup E \, = \, C \cup F \, .
	\end{align}
Let us analyze them label by label. For each color index $i \in \{1, \cdots,N\}$, we record whether it appears in A,B,C,D,E,F.
The above constraints, together with the fact that each determinant minor has distinct row and column labels, allow only the following six patterns:
	\begin{align}
		\varnothing \, , \qquad AB \, , \qquad CD \, , \qquad EF \, , \qquad ACE \, , \qquad BDF \, .
	\end{align}
The first three nonempty patterns are local two-edge patterns and only involve two different types of matrices.  The patterns $ACE$ and $BDF$ are the two opposite orientations of a triangular color-flow pattern.
Let us denote
	\begin{align}
		a \, : = \, \#(AB) \, , \quad b \, := \, \#(CD) \, , \quad c \, := \, \#(EF) \, , \quad s \, := \, \#(ACE) \, , \quad t\, := \, \#(BDF) \, ,
	\end{align}
    where $\#(\cdot)$ represents the number of $\cdot$.
Then, the sizes of the contraction sets are
	\begin{align}
		|A| \, = \, a+s \, , \quad |B| \, = \, a+t, \quad |C| \, = \, b+s \, , \quad |D| \, = \, b+t \, , \quad |E| \, = \, c+s \, , \quad |F| \, = \, c+t \, .
	\end{align}
We also have $|A|=|B|=m$, $|C|=|D|=n$, and $|E|=|F|=l$.  Hence
	\begin{align}
		s \, = \, t \, , \qquad a \, = \, m-t \, , \qquad b \, = \, n-t \, , \qquad c \, = \, l-t \, .
	\end{align}
The total number of distinct color labels used by the contraction pattern is
	\begin{align}
		L \, = \, a+b+c+2t \, = \, m+n+l-t \, .
	\end{align}
For fixed $m,n,\ell,t$, the labels can be chosen in
	\begin{align}
		\frac{N!}{(N-L)! \, (m-t)! \, (n-t)! \, (l-t)! \, (t!)^2}
	\end{align}
ways.  The bijections that pair the endpoints on the three edges give
	\begin{align}
		m! \, n! \, l!
	\end{align}
additional choices.

The relative determinant signs reduce to a simple rule: a pair of opposite triangular patterns $ACE$ and $BDF$ contributes one minus sign.  Thus $t$ such pairs give $(-1)^t$.  Combining label choices, endpoint pairings, propagators, and signs, the coefficient of $x^m y^n z^l$ is
	\begin{align}
		C_N(m,n,l) \, = \, \sum_{\substack{0\le t\le \min(m,n,l)\\ m+n+\ell-t\le N}} (-1)^t \frac{N!}{(N-m-n-l+t)!\,(m-t)!\,(n-t)!\,(l-t)!\,(t!)^2}  \frac{m!\,n!\,l!}{N^{m+n+l}} \, .
	\end{align}
Therefore, the direct Wick-contraction answer is
	\begin{align}
		Z_N(x,y,z) \, := \, Z_{1,1,1}^{(N)}(u,v,w) \, = \, \sum_{m,n,l\ge 0}C_N(m,n,l)x^m y^n z^l \, .
	\end{align}
Equivalently, we set
	\begin{align}
		m \, = \, a+t \, , \qquad n \, = \, b+t \, , \qquad l \, = \, c+t \, .
	\end{align}
Then the exact finite-$N$ polynomial becomes
	\begin{align}
		Z_N(x,y,z) \, = \, \sum_{\substack{a,b,c,t\ge0\\ a+b+c+2t\le N}} (-1)^t \frac{(a+t)!\,(b+t)!\,(c+t)!}{(N-a-b-c-2t)!\,a!\,b!\,c!\,(t!)^2} \frac{N!}{N^{a+b+c+3t}} \, x^{a+t}y^{b+t}z^{c+t} \, .
	\end{align}
We can also check that this exact result reproduces the large $N$ partition function obtained in the F-type model in the previous subsection.
This contribution comes from the term with $a=b=c=0$ and $t=N/2$. Assuming $N$ is an even integer, this contribution is 
	\begin{align}
		Z_N(x,y,z) \, = \, (-1)^{N/2} \, \frac{N!(N/2)!}{N^{3N/2}} \, (G_{12} G_{23} G_{31})^{N/2} \, .
	\end{align}
By using Stirling’s expansion and dropping the overall constant depending on the sources as in section~\ref{sec:f3}, we find
	\begin{align}
		Z_N(x,y,z) \, \approx \, (-1)^{N/2} \sqrt{2} \pi N \left( \frac{\k_1\k_2\k_3 g^{3/2}}{4} \right)^N \, e^{-3N/2} \, .
	\end{align}
This result agrees with the large $N$ partition function up to one-loop combining (\ref{eq:Z_tree-3F}) and (\ref{eq:Z_oneloop-3F}).

Another two simple checks are also useful. If only the $12$ channel is present, then
	\begin{align}
		Z_N(x,0,0) \, = \, \sum_{m=0}^{N}\frac{N!}{(N-m)!}\frac{x^m}{N^m} \, = \, \sum_{m=0}^{N}\frac{(N)_m}{N^m}x^m,
	\end{align}
which is the two-determinant annulus polynomial.  If the intended strict V-type model has no direct $13$ propagator, set $z=0$ and obtain
	\begin{align}
		Z_N(x,y,0) \, = \, \sum_{\substack{a,b\ge0\\ a+b\le N}} \frac{N!}{(N-a-b)!}\frac{x^a y^b}{N^{a+b}} \, .
	\end{align}

\section{Character Expansions}
\label{sec:character}
In this section, we consider the character expansion~\cite{Corley:2001zk,Brown:2010af} of V- and F-type models for three-point functions.\footnote{We use the knowledge of the symmetric group. See, for instance,~\cite{sagan2001symmetric,Macdonald:1995sfa} for the textbooks.}

\subsection{V-type character expansion}
\label{sec:V-character}
In the V-type model, we are interested in the class of operators 
\begin{align}
    \Tr(\alpha_{1}M_{1}^{\otimes k_{1}})\,\Tr(\alpha_{2}M_{2}^{\otimes k_{2}})\,\Tr(\alpha_{3}M_{3}^{\otimes k_{3}})
\end{align}
for $\alpha_{\beta}\in S_{k_{\beta}}$ with $\beta=1,2,3$. The notation $\Tr(\alpha_{\b}M_{\b}^{\otimes k_{\b}})$ is the same as the one introduced in~\cite{Brown:2010af}. Let $[\mu_{1},\mu_{2},\ldots,\mu_{p}]$ be a partition of $k_{\b}=\sum_{i}\mu_{i}$. By using this partition, we compactly write a multitrace operator as
\begin{align}
    \Tr(\alpha_{\b}M_{\b}^{\otimes k_{\b}}):=\Tr M_{\b}^{\mu_{1}}\Tr M_{\b}^{\mu_{2}}\ldots \Tr M_{\b}^{\mu_{p}}\,.
\end{align}
The element $\alpha_{\b}$ is chosen to reproduce the contraction of the indices of matrices for the right-hand side. 
At the level of the correlator, we are interested in 
\begin{align}\label{eq:target-V}
     \left\langle\,\Tr(\alpha_{1}M_{1}^{\otimes k_{1}})\,\Tr(\alpha_{2}M_{2}^{\otimes k_{2}})\,\Tr(\alpha_{3}M_{3}^{\otimes k_{3}})\,\right\rangle_{V}\,,
\end{align}
where the correlator for the product of trace operators $\mathcal{O}$ is defined by
\begin{align}
    \langle\,\mathcal{O}\,\rangle_{V}=\frac{1}{Z_{\mathrm{3pt}}}\int dM_{1}dM_{2}dM_{3\,N\times N}\,\mathcal{O}\,e^{-\frac{N}{g}S}
\end{align}
with
\begin{align}
    S=\frac{1}{2}\Tr \left(\sqrt{Y} M_{\alpha}K_{\alpha\beta}\sqrt{Y}M_{\beta}\right)\,.
\end{align}
In the V-type model, the two-point function is given by
\begin{align}
		\big\la (M_\a)_{ij} (M_\b)_{kl} \big\ra_{V} \, = \, \frac{G_{\a\b}}{N}\, \frac{\delta_{il}\delta_{jk}}{\sqrt{y_i y_j}} \, 
	\end{align}
with 
	\begin{align}
		G_{\a\a} \, = \, 0 \, , \qquad \qquad G_{\a\b} \, = \,  \frac{g}{2}\k_{\alpha}\k_{\beta} \quad (\a \, \ne \b)\,.
	\end{align}
We notice from the Wick contractions  that the above class of correlators~\eqref{eq:target-V} vanishes unless
\begin{align}
     k_{1}=n_{12}+n_{31},\quad k_{2}=n_{12}+n_{23}\,,\quad k_{3}=n_{23}+n_{31}\,,
\end{align}
where $n_{\a\b}$ is the number of the contractions between $M_{\a}$ and $M_{\b}$.
We can solve these to obtain
\begin{align}
      n_{12}=\frac{1}{2}(k_{1}+k_{2}-k_{3})\,,\quad n_{23}=\frac{1}{2}(k_{2}+k_{3}-k_{1})\,,\quad n_{31}=\frac{1}{2}(k_{3}+k_{1}-k_{2})\,.
\end{align}
Since $n_{12},\,n_{23},\,n_{31}$ are non-negative integers, these satisfy
\begin{align}
      k_{1}\leq k_{2}+k_{3}\,,\quad k_{2}\leq k_{3}+k_{1}\,,\quad k_{3}\leq k_{1}+k_{2}\,,
\end{align}
and
\begin{align}
     d=k_{1}+k_{2}+k_{3}\in 2\mathbb{Z}\,.
\end{align}
Now we need to embed $\alpha_{\b}\in S_{k_{\b}}$ for $\b=1,2,3$ into $S_{d}$. When we consider the permutation of $\{\,1,2,\ldots,d\,\}$, we define $\alpha_{\b}$ as acting only on $\{K_{\b-1}+1,\ldots,K_{\b}\}$, where we define $K_{0}=0$ and $K_{\b}=\sum_{\g=1}^{\b}k_{\g}$ for $\b\neq 0$. We denote the block $\{K_{\b-1}+1,\ldots,{K_{\b}}\}$ by $B_{\b}$.
After this embedding, we define
\begin{align}
    \alpha=\alpha_{1}\alpha_{2}\alpha_{3}\,.
\end{align}
We also define the set of the group element $M_{k_{1},k_{2},k_{3}}$. It is defined by
\begin{align}
    &M_{k_{1},k_{2},k_{3}}=\{\tau\in S_{d}|\,\tau^{2}=e,\,\tau(p)\neq p\,, \tau\,\,\mathrm{represent}\nn \\
    &\qquad\qquad\qquad\qquad\mathrm{the\,\,transpositions\,\,between}\,\,B_{\a}\,\,\mathrm{and}\,\,B_{\b} \,\,\mathrm{for\,\,different}\,\,\a\,\,\mathrm{and}\,\,\b\}\,,
\end{align}
where $e$ is the identity element of $S_{d}$. In short, an element of $M_{k_{1},k_{2},k_{3}}$ represents the Wick contractions for given $\{k_1, k_2, k_3\}$.
By using these elements, we can rewrite the correlator as
\begin{equation}
    \begin{split}
    &\left\langle\,\Tr(\alpha_{1} M_{1}^{\otimes k_{1}})\,\Tr(\alpha_{2}M_{2}^{\otimes k_{2}})\,\Tr(\alpha_{3}M_{3}^{\otimes k_{3}})\,\right\rangle_{V}\\
    &=\left(\frac{g}{2N}\right)^{\frac{d}{2}}\kappa_{1}^{k_{1}}\kappa_{2}^{k_{2}}\kappa_{3}^{k_{3}}\sum_{\tau\in\mathcal{M}_{k_{1},k_{2},k_{3}}}\Tr(\alpha\,\tau\,(Y^{-1/2})^{\otimes d})\\
    &=\left(\frac{g}{2N}\right)^{\frac{d}{2}}\kappa_{1}^{k_{1}}\kappa_{2}^{k_{2}}\kappa_{3}^{k_{3}}\sum_{\sigma\in S_{d}}\sum_{\tau\in\mathcal{M}_{k_{1},k_{2},k_{3}}}\Tr(\sigma\,(Y^{-1/2})^{\otimes d})\,\delta(\sigma\alpha\tau)\,,
    \end{split}
\end{equation}
where we introduced the delta function by
\begin{align}
    \delta(g)=\left\{
    \begin{array}{ll}
        1 & g=e \\
        0 & g\neq e
    \end{array}
\right.\,.
\end{align}
We also used the fact that the conjugate and the inversion do not change the cycle type. 
We denote
\begin{align}
    P_{k_{1},k_{2},k_{3}}=\sum_{\tau\in\mathcal{M}_{k_{1},k_{2},k_{3}}}\Tr(\alpha\,\tau\,(Y^{-1/2})^{\otimes d})\,,
\end{align}
and introduce 
\begin{align}
    \Omega_{d}=\sum_{\sigma\in S_{d}}\Tr(\sigma\,(Y^{-1/2})^{\otimes d})\,\sigma\,,\qquad X_{k_{1},k_{2},k_{3}}=\sum_{\tau\in M_{k_{1},k_{2},k_{3}}}\tau\,.
\end{align}
We can then rewrite $P_{k_{1},k_{2},k_{3}}$ as
\begin{align}
    P_{k_{1},k_{2},k_{3}}=\delta(\,\Omega_{d}\,\alpha\, X_{k_{1},k_{2},k_{3}}\,)\,.
\end{align}
We can simplify $X_{k_{1},k_{2},k_{3}}$ by decomposing the space $M_{k_{1},k_{2},k_{3}}$ by considering $H=S_{k_{1}}\times S_{k_{2}}\times S_{k_{3}}\subset S_{d}$, and we have
\begin{align}
    X_{k_{1},k_{2},k_{3}}=\sum_{h\in H/\mathrm{Stab}_{H}(\tau_{0})}\,h\,\tau_{0}\,h^{-1}\,,\label{eq:stab}
\end{align}
where we define $\tau_{0}$ as a reference Wick contraction and
\begin{align}
    \mathrm{Stab}_{H}(\tau_{0})=\{h\in H|\,h\,\tau_{0}\,h^{-1}=\tau_{0}\}\,.
\end{align}
We take $\tau_{0}$ as contracting first $n_{12}$ matrix of $M_{1}$ with the first $n_{12}$ matrix of $M_{2}$, the next $n_{31}$ matrix of $M_{1}$ with the first $n_{31}$ matrix of $M_{3}$, and the rest. The derivation of~\eqref{eq:stab} is presented in Appendix~\ref{sec:details-character}.

It is known that the delta function can be expanded by the character $\chi$ with the representation $R$
\begin{align}
    \delta(g)=\frac{1}{d\,!}\sum_{R\vdash d}\,d_{R}\,\chi_{R}(g)\,,
\end{align}
where $d_{R}$ is the dimensions of the representation $R$. 
We then obtain
\begin{align}
    P_{k_{1},k_{2},k_{3}}=\frac{1}{d\,!}\sum_{R\vdash d}\,d_{R}\,\chi_{R}(\,\Omega_{d}\,\alpha\, X_{k_{1},k_{2},k_{3}}\,)\,.
\end{align}
 We note that $\Omega_{d}$ is a central element. This follows from the fact that the trace is invariant under the same cycle structure. We then have
 \begin{align}
     \chi_{R}(\,\Omega_{d}\,\alpha\, X_{k_{1},k_{2},k_{3}}\,)=\frac{\chi_{R}(\Omega_{d})}{d_{R}}\chi_{R}(\,\alpha\, X_{k_{1},k_{2},k_{3}}\,)\,.
 \end{align}
 The central element $\Omega_{d}$ can be expressed in terms of the character.
 It is also known that, as a consequence of the Schur-Weyl duality, the  characters of $U(N)$ can be written in terms of the character of $S_{d}$ by
 \begin{align}
     \chi_{R}(U)=\frac{1}{d!}\sum_{\sigma\in S_{d}}\chi_{R}(\sigma)\Tr(\sigma U^{\otimes d})\,,
 \end{align}
 where $U$ is an $N\times N$ matrix. The orthogonality relation
 \begin{align}
     &\sum_{\sigma\in S_{n}}\chi_{R}(\sigma)\chi_{S}(\sigma)=n!\delta_{RS}
 \end{align}
 yields
 \begin{align}
     \Tr(\sigma U^{\otimes d})=\sum_{R\vdash d}\chi_{R}(\sigma)\chi_{R}(U)\,.\label{eq:Schur-Weyl}
 \end{align}
 From these relations, we obtain
 \begin{align}
     \chi_{R}(\Omega_{d})=d!\,\chi_{R}(Y^{-\frac{1}{2}})\,.
 \end{align}

 We can simplify $P_{k_{1},k_{2},k_{3}}$ more by using the representation theory. We denote a representation map of $X$ associated with $R$ by $\Gamma_{R}$. We then have
 \begin{align}
     \chi_{R}(\alpha X_{k_{1},k_{2},k_{3}})=\Tr_{V_{R}}(\Gamma_{R}(\alpha)\Gamma_{R}(X_{k_{1},k_{2},k_{3}}))\,.
 \end{align}
 We denote the representation space associated with an irreducible representation $R$ of $S_{d}$ by $V_{R}$. We consider a restriction of such space by $H$, and we can further decompose $V_{R}$ as
 \begin{align}
     V_{R}\cong\bigoplus_{\lambda}W_{\lambda}\otimes M_{\lambda}^{R}\,,
 \end{align}
 where $\lambda=(r_{1},r_{2},r_{3})$ with $r_{\a}$ of the representation $S_{k_{\a}}$ for $\a=1,2,3$,
 \begin{align}
     W_{\lambda}=V_{r_{1}}\otimes V_{r_{2}}\otimes V_{r_{3}}\,,
 \end{align}
 and $r_{\a}$ is a representation of $S_{k_{\a}}$. We note that $W_{\lambda}$ is an irreducible representation of $H$.
 Since the Wick contraction closes by taking the conjugacy class of $H$, we have
 \begin{align}
     [X_{k_{1},k_{2},k_{3}},h]=0
 \end{align}
 and
 \begin{align}
     \Gamma_{R}(X_{k_{1},k_{2},k_{3}})\,\Gamma_{R}(h)=\Gamma_{R}(h)\,\Gamma_{R}(X_{k_{1},k_{2},k_{3}})\label{eq:commutativity}
 \end{align}
 with $h\in H$.
 On the block of $W_{\lambda}\otimes M_{\lambda}$, the subgroup $H$ acts only on the $W_{\lambda}$ factor
 \begin{align}
     \Gamma_{R}(h)|_{W_{\lambda}\otimes M_{\lambda}^{R}} =\Gamma_{\lambda}(h)\otimes I_{M_{\lambda}^{R}}\,.
 \end{align}
We can also write
 \begin{align}
     \Gamma_{R}(X)|_{W_{\lambda}\otimes M_{\lambda}^{R}}=I_{W_{\lambda}}\otimes X_{\lambda}^{R}\,,\label{eq:Schur-decomp}
 \end{align}
 where the detailed explanation is given in Appendix~\ref{sec:details-character}.
 Since
 \begin{align}
     \Gamma_{R}(\alpha X_{k_{1},k_{2},k_{3}})|_{W_{\lambda}\otimes M_{\lambda}^{R}}=\Gamma_{\lambda}(\alpha)\otimes X^{R}_{\lambda}\,,
 \end{align}
 we find
 \begin{align}
     \chi_{R}(\alpha X_{k_{1},k_{2},k_{3}})=\sum_{r_{1},r_{2},r_{3}}\chi_{r_{1}}(\alpha_{1})\chi_{r_{2}}(\alpha_{2})\chi_{r_{3}}(\alpha_{3})\Tr_{M^{R}_{r_1,r_2,r_3}}(X^{R}_{r_{1},r_{2},r_{3}})\,.
 \end{align}
 Therefore, we have
 \begin{align}
     P_{k_{1},k_{2},k_{3}}=\sum_{R\vdash d}\sum_{r_{1},r_{2},r_{3}}\chi_{R}(Y^{-\frac{1}{2}}) \,\chi_{r_{1}}(\alpha_{1})\chi_{r_{2}}(\alpha_{2})\chi_{r_{3}}(\alpha_{3})\Tr_{M^{R}_{r_1,r_2,r_3}}(X^{R}_{r_{1},r_{2},r_{3}})\,.
 \end{align}

 Let us evaluate the V-type model~\eqref{eq:V-type} in terms of the character expansion. We can expand
 \begin{align}
     e^{-\sum_{k\geq 1}\frac{t_{k}}{k}\Tr M_{1}^{k}}
     &=\sum_{k=0}^{\infty}\sum_{\mu\vdash k}\prod_{p=1}^{k}\frac{(-1)^{i_{p}(\mu)}}{i_{p}(\mu)!p^{i_{p}(\mu)}}[\Tr U^{-p}]^{i_{p}(\mu)}[\Tr M_{1}^{p}]^{i_{p}(\mu)}\nn\\
     &=\sum_{k=0}^{\infty}\sum_{\mu\vdash k}\frac{|[\mu]|}{k!}(-1)^{\sum_{p=1}^{k}i_{p}(\mu)}\Tr(\alpha_{\m}(U^{-1})^{\otimes k})\Tr(\alpha_{\m}(M_{1})^{\otimes k})\,,
 \end{align}
where $i_{p}(\mu)$ is the number of $p$-cycle which is contained in $\alpha_{\mu}$, $|[\mu]|$ is the size of the conjugacy class given by
\begin{equation}
    |[\mu]|=\frac{k!}{\prod_{p=1}^{k}p^{i_{p}(\mu)}i_{p}(\mu)!}\,.
\end{equation}
Note that
\begin{equation}
    \mathrm{sgn}(\alpha_{\mu})=(-1)^{k-l(\mu)}\,,
\end{equation}
where we introduced the number of  parts of the partition by
\begin{align}
    l(\m)=\sum_{p=1}^{k}i_{p}(\mu)\,.
\end{align}

 We then have
 \begin{equation}
     \begin{split}
         &\left\langle\,e^{-\sum_{k_{1}\geq 1}\frac{t_{1k_{1}}}{k_{1}}\Tr M_{1}^{k_{1}}-\sum_{k_{2}\geq 1}\frac{t_{2k_{2}}}{k_{2}}\Tr M_{2}^{k_{2}}-\sum_{k_{3}\geq 1}\frac{t_{3k_{3}}}{k_{3}}\Tr M_{3}^{k_{3}}}\,\right\rangle_{V}\\
         &=\sum_{k_{1},k_{2},k_{3}}\sum_{\mu\vdash k_{1}}\sum_{\nu\vdash k_{2}}\sum_{\rho\vdash k_{3}}\frac{|[\mu]||[\nu]||[\rho]|}{k_{1}!k_{2}!k_{3}!}\Tr(\alpha_{1,\mu}(U^{-1})^{\otimes k_{1}})\Tr(\alpha_{2,\nu}(V^{-1})^{\otimes k_{2}})\Tr(\alpha_{3,\rho}(W^{-1})^{\otimes k_{3}})\\
         &\times \,(-1)^{k_{1}+k_{2}+k_{3}}\sgn(\alpha_{1,\mu})\sgn(\alpha_{2,\n})\sgn(\alpha_{3,\r})\,\left\langle\,\Tr(\alpha_{1,\mu}\,M_{1}^{\otimes k_{1}})\Tr(\alpha_{2,\nu}\,M_{2}^{\otimes k_{2}})\Tr(\alpha_{3,\rho}\,M_{3}^{\otimes k_{3}})\,\right\rangle\\
         &=\sum_{k_{1},k_{2},k_{3}}\sum_{R\vdash d}\sum_{r_{1},r_{2},r_{3}}\kappa_{1}^{k_{1}}\kappa_{2}^{k_{2}}\kappa_{3}^{k_{3}}\left(\frac{g}{2N}\right)^{\frac{d}{2}}\,\\
         &\quad\times\chi_{R}(Y^{-1/2})(-1)^{k_{1}+k_{2}+k_{3}}\chi_{r_{1}}(U^{-1})\chi_{r_{2}}(V^{-1})\chi_{r_{3}}(W^{-1})\Tr_{M_{r^{t}_{1},r^{t}_{2},r^{t}_{3}}^{R}}(X^{R}_{r^{t}_{1},r^{t}_{2},r^{t}_{3}})\,,
     \end{split}
 \end{equation}
 where $r^{t}$ is the transpose of the irreducible representation $r$, and we used the fact
 \begin{align}
      \chi_{R^{t}}(\sigma)=\mathrm{sgn}(\sigma)\,\chi_{R}(\sigma)\,.\label{eq:dual-tableaux}
 \end{align}
We note that, due to the Wick contraction, $d$ should be an even number. We then replace $d$ with $2k$ to obtain
\begin{equation}
     \begin{split} 
         &\left\langle\,e^{-\sum_{k_{1}\geq 1}\frac{t_{1k_{1}}}{k_{1}}\Tr M_{1}^{k_{1}}-\sum_{k_{2}\geq 1}\frac{t_{2k_{2}}}{k_{2}}\Tr M_{2}^{k_{2}}-\sum_{k_{3}\geq 1}\frac{t_{3k_{3}}}{k_{3}}\Tr M_{3}^{k_{3}}}\,\right\rangle_{V}\\
         &=\sum_{\substack{k_{1},k_{2},k_{3}\\k_{1}+k_{2}+k_{3}=2k}}\sum_{R\vdash 2k}\sum_{r_{1},r_{2},r_{3}}\left(\frac{g}{2N}\right)^{k}\kappa_{1}^{k_{1}}\kappa_{2}^{k_{2}}\kappa_{3}^{k_{3}}\\
         &\quad\times\chi_{R}(Y^{-1/2})\chi_{r_{1}}(U^{-1})\chi_{r_{2}}(V^{-1})\chi_{r_{3}}(W^{-1})\Tr_{M_{r_{1}^{t},r_{2}^{t},r_{3}^{t}}^{R}}(X^{R}_{r_{1}^{t},r_{2}^{t},r_{3}^{t}})\,. \label{eq:Vtype-character}
     \end{split}
 \end{equation} 
 We note that the irreducible representation of $S_{d}$ is associated with the Young diagram. Therefore, we  identify the partition of $d$ with the Young diagram. When the number of rows of the Young diagram $l(R)$ satisfies
 \begin{align}
     l(R)> N\,,
 \end{align}
 the character vanishes. 

\subsection{F-type character expansion}
In this subsection, we consider the character expansion of the F-type model of the reduction for the three-point functions.
For this purpose, we rewrite
\begin{align}
    Z_{\mathrm{3F}}=\left\langle\,\exp\left[-\sum_{k=1}^{\infty}\frac{1}{k}\left(-\frac{i}{2}\right)^{k}s_{\frac{k}{2}}\Tr \Lambda^{k}\right]\,\right\rangle_{F}\,,
\end{align}
where the bracket $\langle\,\cdot\,\rangle_{F}$ is the correlation function evaluated by the action
\begin{align}
    S_{0}=\frac{N}{2g}\,\Tr (\Sigma_{1}^{\dagger}\Sigma_{1}+\Sigma_{2}^{\dagger}\Sigma_{2}+\Sigma_{3}^{\dagger}\Sigma_{3})\,.
\end{align}
The two-point functions measured by the action $S_{0}$ are given by
\begin{align}
    \langle\,\Sigma_{1\,ab'}\,\Sigma_{1\,a'b}^{\dagger}\,\rangle=\frac{2g}{N}\,\delta_{ab}\,\delta_{a'b'}\,,\quad
    \langle\,\Sigma_{2\,a'\hat{b}}\,\Sigma_{2\,\hat{a}b'}^{\dagger}\,\rangle=\frac{2g}{N}\,\delta_{a'b'}\,\delta_{\hat{a}\hat{b}}\,,\quad
    \langle\,\Sigma_{3\,\hat{a}b}\,\Sigma_{3\,a\hat{b}}^{\dagger}\,\rangle=\frac{2g}{N}\,\delta_{ab}\,\delta_{\hat{a}\hat{b}}\,.\label{eq:Wick-F-type}
\end{align}
Similar to the analysis of section~\ref{sec:V-character}, we have
\begin{equation}
    \begin{split}
        &\exp\left[-\sum_{k=1}^{\infty}\frac{1}{k}\left(-\frac{i}{2}\right)^{k}s_{\frac{k}{2}}\Tr\Lambda^{k}\right]\\
        &=\sum_{k=0}^{\infty}\sum_{\mu\vdash k}\prod_{p=1}^{k}\frac{\left[-\frac{1}{p}(-\frac{i}{2})^{p}\Tr Y^{-\frac{p}{2}}\Tr\Lambda^{p}\right]^{i_{p}(\mu)}}{i_{p}(\mu)!}\\
        &=\sum_{k=0}^{\infty}\sum_{\mu\vdash k}\,(-1)^{\sum_{p}(i_{p}(\mu)+pi_{p}(\mu))}\,\left(\frac{i}{2}\right)^{\sum_{p}pi_{p}(\mu)}\,\frac{|[\mu]|}{k!}\,\Tr[\alpha_{\mu}(Y^{-\frac{1}{2}})^{\otimes k}]\,\Tr(\alpha_{\mu}\Lambda^{\otimes k})\\
        &=\sum_{k=0}^{\infty}\sum_{\mu\vdash k}\,\mathrm{sgn}(\alpha_{\mu})\,\left(\frac{i}{2}\right)^{k}\,\frac{|[\mu]|}{k!}\,\Tr[\alpha_{\mu}(Y^{-\frac{1}{2}})^{\otimes k}]\,\Tr(\alpha_{\mu}\Lambda^{\otimes k})\,.
    \end{split}
\end{equation}
Since  the number of $\Sigma$'s and $\Sigma^{\dagger}$'s is equal due to the Wick contraction rule~\eqref{eq:Wick-F-type}, the power of $\Lambda$ thus becomes even. We then have 
\begin{equation}
    \begin{split}
        &\left\langle\,\exp\left[-\sum_{k=1}^{\infty}\frac{1}{k}\left(-\frac{i}{2}\right)^{k}s_{\frac{k}{2}}\Tr\Lambda^{k}\right]\,\right\rangle_{F}\\
        &=\sum_{k=0}^{\infty}\sum_{\mu\vdash 2k}\,\mathrm{sgn}(\alpha_{\mu})\,\frac{(-1)^{k}}{2^{2k}}\,\frac{|[\mu]|}{(2k)!}\,\Tr[\alpha_{\mu}(Y^{-\frac{1}{2}})^{\otimes 2k}]\,\langle\,\Tr(\alpha_{\mu}\Lambda^{\otimes 2k})\,\rangle\,.
    \end{split}
\end{equation}

For further analysis, we introduce
\begin{equation}
    \begin{split}
        &\widetilde{X}_{12}=U^{-1}\Sigma_{1}\,,\qquad \widetilde{X}_{13}=U^{-1}\Sigma_{3}^{\dagger}\,,\qquad
        \widetilde{X}_{21}=V^{-1}\Sigma_{1}^{\dagger}\,,\\
        &\widetilde{X}_{23}=V^{-1}\Sigma_{2}\,,\qquad
        \widetilde{X}_{31}=W^{-1}\Sigma_{3}\,,\qquad \widetilde{X}_{32}=W^{-1}\Sigma_{2}^{\dagger}\,,
    \end{split}
\end{equation}
and consider the set $W_{2k}(\a_{\m})$ whose element is described by
\begin{align}
    w=\Tr \a_{\m}(X_{1}\otimes X_{2}\otimes \ldots\otimes X_{2k})
\end{align}
with $X_{j}\in\{\widetilde{X}_{12},\widetilde{X}_{13},\widetilde{X}_{21},\widetilde{X}_{23},\widetilde{X}_{31},\widetilde{X}_{32}\}$.
To define $w$ and $W_{2k}(\a_{\m})$ appropriately, we need the condition of the consistency of the matrix products and $\a_{\m}$. To describe this requirement, let us consider the trace
\begin{align}
    \Tr(\a_{\m}(A_{1}\otimes\ldots\otimes A_{2k}))=\sum_{a_{1},\ldots,a_{2k}}\prod_{p=1}^{2k}(A_{p})_{a_{p}a_{\a_{\m}(p)}}\,.
\end{align}
 We define a slot $p$ by the $p$-th tensor element. We also denote the set of slots by
\begin{align}
    [2k]=\{1,2,\ldots,2k\}\,.
\end{align}
We introduce 
\begin{align}
    \widetilde{w}:[2k]\rightarrow\{12,13,21,23,31,32\}\,.
\end{align}
The map $\widetilde w(p)=\a\b$ represent that the slot $p$ is occupied by $X_{\a\b}$ with $\a,\b=1,2,3$ and $\a\neq\b$. We also define
\begin{align}
    s(p)=\a\,,\qquad t(p)=\b\,,
\end{align}
when $\widetilde{w}=\a\b$. One condition $w\in W_{2k}$, which we call closed walk condition, is described by
\begin{align}
    s(\a_{\m}(p))=t(p)\,.
\end{align}
We denote $n_{i}(w)$ and $\overline{n}_{i}(w)$ by the number of the appearance of $\Sigma_{i}$ and $\Sigma_{i}^{\dagger}$, respectively. We note that these should satisfy the relation 
\begin{align}
    n_{i}(w)=\overline{n}_{i}(w)\label{eq:balance}
\end{align}
for the non-vanishing correlator
due to the Wick contraction rule~\eqref{eq:Wick-F-type}.
We also demand the condition~\eqref{eq:balance} for $w\in W_{2k}$.

The contraction pattern is described by a permutation. The contraction between $\Sigma_{i}$ and $\Sigma_{i}^{\dagger}$ can be characterized by
\begin{align}
    \gamma^{(i)}\in S_{n_i(w)}\,.
\end{align}
Combined with the trace structure and the Wick contraction patterns, the closed cycle of indices can be captured by three kinds of index lines associated with the index label $a$, $a'$, and $\hat{a}$. We denote these index lines by $\pi_{Q}(w;\alpha,\gamma)\in S_{n_{1}(w)+n_{3}(w)}$, $\pi_{R}(w;\alpha,\gamma)\in S_{n_{1}(w)+n_{2}(w)}$, and $\pi_{S}(w;\alpha,\gamma)\in S_{n_{2}(w)+n_{3}(w)}$, respectively, where $\alpha$ is the original trace structure, and $\gamma=(\gamma^{(1)},\gamma^{(2)},\gamma^{(3)})$.
We have
\begin{equation}
    \begin{split}
        &\left\langle\,\exp\left[-\sum_{k=1}^{\infty}\frac{1}{k}\left(-\frac{i}{2}\right)^{k}s_{\frac{k}{2}}\Tr\Lambda^{k}\right]\,\right\rangle_{F}\\
        &=\sum_{k=0}^{\infty}\sum_{\mu\vdash 2k}\sum_{w\in W_{2k}}\sum_{\g}(-1)^{k}\mathrm{sgn}(\alpha_{\mu})\,\frac{|[\mu]|}{(2k)!}\,\kappa_{1}^{n_{1}(w)+n_{3}(w)}\kappa_{2}^{n_{1}(w)+n_{2}(w)}\kappa_{3}^{n_{2}(w)+n_{3}(w)}\\
        &\times\left(\frac{g}{2N}\right)^{k}\Tr[\alpha_{\mu}(Y^{-\frac{1}{2}})^{\otimes 2k}]\Tr\pi_{Q}(w;\a_{\m},\g)(U^{-1})^{\otimes (n_{1}(w)+n_{3}(w))}\\
        &\times \Tr\pi_{R}(w;\a_{\m},\g)(V^{-1})^{\otimes (n_{1}(w)+n_{2}(w))}\Tr\pi_{S}(w;\a_{\mu},\g)(W^{-1})^{\otimes (n_{2}(w)+n_{3}(w))}\,.
    \end{split}
\end{equation}
By using~\eqref{eq:Schur-Weyl}, we have 
\begin{equation}
    \begin{split}
        &\left\langle\,\exp\left[-\sum_{k=1}^{\infty}\frac{1}{k}\left(-\frac{i}{2}\right)^{k}s_{\frac{k}{2}}\Tr\Lambda^{k}\right]\,\right\rangle_{F}\\
        &=\sum_{k=0}^{\infty}\sum_{\mu\vdash 2k}\sum_{w\in W_{2k}}\sum_{\g}\frac{|[\mu]|}{(2k)!}(-1)^{k}\mathrm{sgn}(\alpha_{\mu})\left(\frac{g}{2N}\right)^{k}\kappa_{1}^{n_{1}(w)+n_{3}(w)}\kappa_{2}^{n_{1}(w)+n_{2}(w)}\kappa_{3}^{n_{2}(w)+n_{3}(w)}\\
        &\times \sum_{\substack{\mathcal{R}\vdash 2k}}\sum_{\substack{R_{Q}\vdash n_{1}+n_{3}}}\,\sum_{\substack{R_{R}\vdash n_{1}+n_{2}}}\sum_{\substack{R_{S}\vdash n_{2}+n_{3}}} \chi_{\mathcal{R}}(Y^{-\frac{1}{2}}) \chi_{R_{Q}}(U^{-1})\chi_{R_{R}}(V^{-1})\chi_{R_{S}}(W^{-1})\\
        &\times \chi_{\mathcal{R}}(\alpha_{\mu}) \chi_{R_{Q}}(\pi_{Q}(w;\a_{\m},\g))\chi_{R_{R}}(\pi_{R}(w;\a_{\m},\g))\chi_{R_{S}}(\pi_{S}(w;\a_{\m},\g))\,.\label{eq:Ftype-character}
    \end{split}
\end{equation}

\subsection{Equivalence of the two open string descriptions}
In this subsection, we compare the character expansions of two open string descriptions~\eqref{eq:Vtype-character} and~\eqref{eq:Ftype-character}. To see this equivalence, let us fix $(k,k_{1},k_{2},k_{3})$ in the V-type model and $(k,n_{1}(w),n_{2}(w),n_{3}(w))$ in the F-type model. We denote the subset of $W_{2k}(\a_{\m})$ with this fixing by $W^{(k)}_{k_{1},k_{2},k_{3}}(\a_{\m})$. The identification
\begin{align}
    k^{(V)}=k^{(F)}\,,\quad k_{1}=n_{1}+n_{3}\,,\quad k_{2}=n_{1}+n_{2}\,,\quad k_{3}=n_{2}+n_{3}\,,\label{eq:VF-fixing1}
\end{align}
where $k^{(V)}$ is $k$ in~\eqref{eq:Vtype-character} and $k^{(F)}$ is $k$ in~\eqref{eq:Ftype-character}, and we  simply write $n_{i}$ instead of $n_{i}(w)$ for simplicity. Then, part of the coefficient matches:
\begin{align}
    \left(\frac{g}{2N}\right)^{k^{(V)}}\kappa_{1}^{k_{1}}\kappa_{2}^{k_{2}}\kappa_{3}^{k_{3}}=\left(\frac{g}{2N}\right)^{k^{(F)}}\kappa_{1}^{n_{1}+n_{3}}\kappa_{2}^{n_{1}+n_{2}}\kappa_{3}^{n_{2}+n_{3}}\,.
\end{align}
For a fixed representation, the identification
\begin{align}
    R=\mathcal{R}\,,\quad r_{1}=R_{Q}\,,\quad r_{2}=R_{R}\,,\quad r_{3}=R_{S} \label{eq:VF-fixing2}
\end{align}
yields the matching
\begin{align}
    \chi_{\mathcal{R}}(Y^{-\frac{1}{2}})\chi_{r_{1}}(U^{-1})\chi_{r_{2}}(V^{-1})\chi_{r_{3}}(W^{-1})=\chi_{R}(Y^{-\frac{1}{2}})\chi_{R_{Q}}(U^{-1})\chi_{R_{R}}(V^{-1})\chi_{R_{S}}(W^{-1})\,.
\end{align}
Therefore, to see the equivalence, we have to show
\begin{align}
    C_{\mathrm{V-type}}=C_{\mathrm{F-type}}\,,
\end{align}
where
\begin{align}
    C_{\mathrm{V-type}}&=\Tr_{M^{R}_{r_{1}^{t},r_{2}^{t},r_{3}^{t}}}(X^{R}_{r_{1}^{t},r_{2}^{t},r_{3}^{t}})\,,\\
    C_{\mathrm{F-type}}&=\sum_{\mu\vdash 2k}\sum_{w\in W^{(k)}_{k_{1},k_{2},k_{3}}(\a_{\m})}\sum_{\g}\frac{|[\mu]|}{(2k)!}(-1)^{k}\sgn(\alpha_{\mu})\chi_{R}(\alpha_{\mu})\nn\\
    &\qquad \times\chi_{R_{Q}}(\pi_{Q}(w;\alpha_{\mu},\g))\chi_{R_{R}}(\pi_{R}(w;\alpha_{\mu},\g))\chi_{R_{S}}(\pi_{S}(w;\alpha_{\mu},\g))\,.
\end{align}

Let us first deform the V-type coefficient $C_{\mathrm{V-type}}$. Since
\begin{align}
    \chi_{R}(\alpha X_{k_{1},k_{2},k_{3}})=\sum_{\substack{r_{\b}\vdash k_{\b}\\\b=1,2,3}} \chi_{r_{1}^{t}}(\alpha_{1})\chi_{r_{2}^{t}}(\alpha_{2})\chi_{r_{3}^{t}}(\alpha_{3}) \Tr_{M_{r_{1}^{t},r_{2}^{t},r_{3}^{t}}^{R}} (X_{r_{1}^{t},r_{2}^{t},r_{3}^{t}}^{R})\,,
\end{align}
the trace can be rewritten as
\begin{align}
    C_{\mathrm{V-type}}=\frac{1}{k_{1}!k_{2}!k_{3}!}\sum_{\substack{\alpha_{\b}\in S_{k_{\b}}\\\b=1,2,3}}\sum_{\tau\in M_{k_{1},k_{2},k_{3}}}\chi_{r_{1}^{t}}(\alpha_{1})\chi_{r_{2}^{t}}(\alpha_{2})\chi_{r_{3}^{t}}(\alpha_{3}) \chi_{R}(\alpha\tau)\,.
\end{align}
The F-type coefficient can also be simplified. We take the sum over the whole $S_{2k}$ instead of its conjugacy class to obtain
\begin{align}
    C_{\mathrm{F-type}}&=\frac{(-1)^{k}}{(2k)!}\sum_{\sigma\in S_{2k}}\sum_{w\in W^{(k)}_{k_{1},k_{2},k_{3}}(\s)}\sum_{\g}\sgn(\sigma)\nn\\
    &\qquad\times\chi_{R}(\sigma)\chi_{R_{Q}}(\pi_{Q}(w;\s,\g))\chi_{R_{R}}(\pi_{R}(w;\s,\g))\chi_{R_{S}}(\pi_{S}(w;\s,\g))\,.
\end{align}

Let us construct the one-to-one correspondence between the V- and F-type models. We have to match the V-type data
\begin{align}
    \mathcal{V}:=({\bm B}:=(B_{1},B_{2},B_{3}),(\a_{1},\a_{2},\a_{3}),\tau)
\end{align}
with the F-type data
\begin{align}
    \mathcal{F}:=(\sigma,w,\gamma)
\end{align}
under the parameter fixing~\eqref{eq:VF-fixing1} and~\eqref{eq:VF-fixing2}. 
We define the set of slots
\begin{align}
    B_{\a}=\{p\in[2k]|s(p)=\a\}\,.
\end{align}
The block $B_{1},\,B_{2}$ and $B_{3}$ contains $U^{-1},\,V^{-1}$ and $W^{-1}$, respectively. The number of elements of $B_{\a}$ is given by $k_{\a}$. We note that $\widetilde{w}(p)=\a\b$ can only be contracted with $\widetilde{w}(q)=\b\a$. Therefore, the contraction is taken between different blocks. We then introduce the contraction map $\tau$ by
\begin{align}
    \tau(p;\g)=q\,,\qquad \tau(q;\g)=p\,.
\end{align}
This definition for $\tau$ satisfies
\begin{align}
    \tau^{2}=e\,,\qquad \tau(p)\neq p\,.
\end{align}
Moreover, 
\begin{align}
    \widetilde{w}(\tau(p))=t(p)s(p)\,.
\end{align}

Let us then explicitly match the data $\mathcal{V}$ and $\mathcal{F}$. Let us start from $\mathcal{F}$ to  $\mathcal{V}$. We can construct ${\bm B}$ by defining
\begin{align}
    B_{\a}=s^{-1}(\alpha)\,.
\end{align}
We then put 
\begin{align}
    \a=\s\t
\end{align}
to obtain the element $\a\in H$. 

Let us next consider reconstructing  $\mathcal{V}$ to $\mathcal{F}$. We put
\begin{align}
    \s=\a\t\,,
\end{align}
and if $p\in B_{\a}\,,\tau(p)\in B_{\b}$, we define
\begin{align}
    \widetilde{w}(p)=\a\b\,.
\end{align}
We then construct $\g$ to reproduce the Wick contraction for each $\{p,\tau(p)\}$. Let us check the closed walk condition. When the slot $p$ is in $B_{\a}$ and the slot $\tau(p)$ is in $B_{\b}$,
\begin{align}
    t(p)=\b\,.
\end{align}
Since $\s(p)=\a\t(p)$ with $\t(p)\in B_{\b}$ and $\a$ preserves the block, we have $\s(p)\in B_{\b}$ and satisfies the closed walk condition
\begin{align}
    s(\s(p))=\b=t(p)\,.
\end{align}

Let us finally present that the above correspondence is one-to-one. We denote $\Phi:\mathcal{F}\rightarrow\mathcal{V}$ and $\Psi:\mathcal{V}\rightarrow\mathcal{F}$. Let us first consider $\s'$, which is constructed from applying $\Psi$ after $\Phi$. From the property $\tau^{2}=e$, we have
\begin{align}
    \s'=(\s\t)\t=\s\,.
\end{align}
The closed walk $w$ and Wick contraction $\g$ are recovered by construction. Let us next consider $\a'$, which is constructed from applying $\Phi$ after $\Psi$. We have
\begin{align}
    \a'=(\a\t)\t=\a\,.
\end{align}
The block ${\bm B}$ and the contraction $\t$ are recovered by construction. We then find
\begin{align}
    \Phi\circ\Psi=\Psi\circ\Phi=\mathrm{id}
\end{align}
and the correspondence is one-to-one. Note that when we choose the block from the slots, there are $\frac{(2k)!}{k_{1}!k_{2}!k_{3}!}$ choices.

We can deform the F-type coefficient to
\begin{align}
    C_{\mathrm{F-type}}&=\frac{(-1)^{k}}{k_{1}!k_{2}!k_{3}!}\sum_{\substack{\a_{\b}\in S_{k_{\b}}\\\b=1,2,3}}\sum_{\t\in M_{k_{1},k_{2},k_{3}}}\sgn(\a\t)\chi_{R}(\a\t)
\chi_{R_{Q}}(\a_{1})\chi_{R_{R}}(\a_{2})\chi_{R_{S}}(\a_{3})\,,
\end{align}
where $\a$ is defined from $\a_{1},\,\a_{2}$ and $\a_{3}$ as before.
Since the sign of $\t$ is given by the number of transpositions, which is equal to the number of the Wick contraction, we have
\begin{align}
    \sgn(\a\t)=\sgn(\a)\sgn(\t)=(-1)^{k}\sgn(\a_{1})\sgn(\a_{2})\sgn(\a_{3})\,.
\end{align}
We then obtain
\begin{align}
    C_{\mathrm{F-type}}&=\frac{1}{k_{1}!k_{2}!k_{3}!}\sum_{\substack{\a_{\b}\in S_{k_{\b}}\\\b=1,2,3}}\sum_{\t\in M_{k_{1},k_{2},k_{3}}}\chi_{R}(\a\t)
\chi_{R_{Q}^{t}}(\a_{1})\chi_{R_{R}^{t}}(\a_{2})\chi_{R_{S}^{t}}(\a_{3})\,.
\end{align}
This is exactly the same as $C_{\mathrm{V-type}}$ under~\eqref{eq:VF-fixing1} and~\eqref{eq:VF-fixing2}.

\subsection{Reduction to single giant}
In this subsection, we consider the character expansion for the single giant case. In section~\ref{sec:single-twopt}, we carry out the character expansion for the two-point functions. In section~\ref{sec:single-threept}, we discuss the reduction to single giant by using the result of the previous section.

\subsubsection{Two-point functions}\label{sec:single-twopt}
We consider the single giant partition function for the V-type two-matrix model, which is given by
\begin{equation}
    Z^{(N)}_{1,1}=\frac{\det(NY)^{N}}{(2\pi g)^{N^{2}}}\int dK dM e^{-\frac{N}{g}\mathrm{tr}(\sqrt{Y}K\sqrt{Y}M)-\sum_{k\geq 1}\frac{t_{k}}{k}\mathrm{tr}(K^{k})-\sum_{k\geq 1}\frac{\overline{t}_{k}}{k}\mathrm{tr}(M^{k})}\,.\label{eq:V-2-matrix}
\end{equation}
We note that the sign before the Miwa variables is different compared to~\cite{Brown:2010af}.

Let us consider the character expansion of~\eqref{eq:V-2-matrix}.
By using the technique explained in section~\ref{sec:V-character}, we have
\begin{align}
    \exp\left(-\sum_{k\geq1}\frac{1}{k}t_{k}\,\mathrm{tr} M^{k}\right)
        =\sum_{k=0}^{\infty}\sum_{R\vdash k}(-1)^{k}\chi_{R}(X^{-1})\chi_{R^{t}}(M)\,.
\end{align}
We can similarly calculate the terms involving $K$ and we obtain
\begin{align}
    Z_{1,1}^{(N)} =\sum_{k_{1},k_{2}=0}^{\infty}\sum_{R\vdash k_{1}}\sum_{T\vdash k_{2}}(-1)^{k_{1}+k_{2}}\chi_{R}(X^{-1})\chi_{T}(V^{-1})\langle\,\chi_{R^{t}}(M)\,\chi_{T^{t}}(K)\,\rangle_{K,M}\,,\label{eq:single-giant-2mat-temp}
\end{align}
where
\begin{align}
    \langle\,\mathcal{O}(K,M)\rangle_{K,M}=\frac{\det(NY)^{N}}{(2\pi g)^{N^{2}}}\int dK dM \mathcal{O}(K,M)\, e^{-\frac{N}{g}\mathrm{tr}(\sqrt{Y}K\sqrt{Y}M)}\,.
\end{align}
By following the discussion described in~\cite{Brown:2010af}, we find
\begin{align}
    \langle\,\chi_{R}(M)\chi_{T}(K)\,\rangle_{K,M}=\frac{k_{1}!}{d_{R}}\chi_{R}\left(\frac{g}{N}Y^{-1}\right)\delta_{k_{1},k_{2}}\delta_{R,T}
\end{align}
We then have
\begin{align}
    Z_{1,1}^{(N)}=\sum_{k=0}^{\infty}\sum_{R\vdash k}\frac{k!}{d_{R}}\chi_{R}(X^{-1})\chi_{R}(V^{-1})\chi_{R^{t}}\left(\frac{g}{N}Y^{-1}\right)\,.
\end{align}

For a single giant,
\begin{align}
    \chi_{R}(X^{-1})=0
\end{align}
unless
\begin{align}
    R=[k]\,.
\end{align}
This representation is the trivial representation of $S_{k}$.
For $U(N)$ character, 
the transposed tableaux
\begin{align}
    R^{t}=[1^{k}]
\end{align}
are involved,
and this is the sign representation of $S_{k}$.
We then have
\begin{align}
    \chi_{[k]}(x^{-1})=x^{-k}\,,\qquad \chi_{[k]}(v^{-1})=v^{-k}\,.
\end{align}
Since
\begin{align}
    d_{[k]}=1\,,
\end{align}
we have
\begin{equation}
    Z_{1,1}^{(N)} =\sum_{k=0}^{\infty}k!\left(\frac{1}{xv}\right)^{k}\chi_{[1^{k}]}\left(\frac{g}{N}Y^{-1}\right)\,.
\end{equation}
Since the representation $R^{t}=(1^{k})$ is the sign representation, we have
\begin{equation}
    \chi_{[1^{k}]}\left(\frac{g}{N}Y^{-1}\right)=\left(\frac{g}{N}\right)^{k}\sum_{\substack{I\subset\{1,\ldots,N\}\\|I|=k}}\prod_{i\in I}\frac{1}{y_{i}}\,.
\end{equation}
For the uniform source
\begin{align}
    Y=yI_{N}\,,
\end{align}
we  reproduce the result obtained in section~\ref{sec:f2} and~\ref{sec:v2}:
\begin{align}
		Z_{1,1}^{(N)}(x,v) \, = \, \sum_{r=0}^{N} \frac{(N)_k}{N^k} \, q_y^k \, .
	\end{align}

\subsubsection{Three-point functions}
\label{sec:single-threept}
Now we explain how the exact single-giant finite-$N$ polynomial for the V-type three-point function obtained in section~\ref{sec:v3} is reproduced from the three-point character expansion by restricting to one determinant at each external point, i.e. $Q=R=S=1$. 

We start from the V-type character expansion written in the common positive-alphabet basis. 
The closed-line matrix is denoted by $F$.  In applications to the V/F duality, we take $F=Y^{-1/2}$, and for the finite-$N$ polynomial result, we later set $F=I_N$.
The character expansion is written as
	\begin{align}
		Z_{\mathrm V} \, &= \, \sum_{\substack{k_{1},k_{2},k_{3}\ge0\\k_{1}+k_{2}+k_{3}=2k}} \left(\frac{g}{2N}\right)^k \kappa_1^{k_1}\kappa_2^{k_2}\kappa_3^{k_3} \\
		&\quad \times \sum_{\substack{\mathcal{R}\vdash d\\ \ell(\mathcal{R})\le N}} \sum_{R_Q\vdash k_1} \sum_{R_R\vdash k_2} \sum_{R_S\vdash k_3}
		\chi_{\mathcal{R}}(F) \chi_{R_Q}(U^{-1}) \chi_{R_R}(V^{-1}) \chi_{R_S}(W^{-1}) \mathcal{V}_{\mathcal{R};R_Q^t,R_R^t,R_S^t} \, , \nn
	\end{align}
with
\begin{align}
    \pi=\pi_{1}\pi_{2}\pi_{3}\in H
\end{align}
and
	\begin{align}
		\mathcal{V}_{\mathcal{R};R_Q^t,R_R^t,R_S^t} \, &= \, \frac{1}{k_{1}!k_{2}!k_{3}!} \sum_{\pi_1\in S_{k_1}} \sum_{\pi_2\in S_{k_2}} \sum_{\pi_3\in S_{k_3}}
		\chi_{R_Q^t}(\pi_1) \chi_{R_R^t}(\pi_2) \chi_{R_S^t}(\pi_3) \sum_{\tau\in M_{k_1,k_2,k_3}} \chi_{\mathcal{R}}(\pi\tau) \, .
	\end{align}	
Here $R_Q,R_R,R_S$ label the three open sectors, and the transpose appears because the determinant source is a negative plethystic alphabet on the V-type side.  The expansion is already written in terms of ordinary positive alphabets $U^{-1},V^{-1},W^{-1}$.

Now we restrict to one determinant at each external point:
	\begin{align}
		Q=R=S=1 \, , \qquad U=u \, , \quad V=v \, ,\quad W=w \, .
	\end{align}
Thus the open matrices are one-dimensional.  Since we are considering a single giant, a Schur character in one variable vanishes unless the Young diagram has at most one row as we discussed in section~\ref{sec:single-twopt}:
	\begin{align}
		\chi_\lambda(u^{-1})=0 \quad\text{unless}\quad \lambda=[L] \quad\text{for some}\ L\ge0 \, .
	\end{align}
For the surviving one-row representation, we have
	\begin{align}
		\chi_{[L]}(u^{-1}) \, = \, u^{-L} \, .
	\end{align}
Therefore, the three open representation sums in the character expansion collapse to
	\begin{align}
		R_Q=[k_1], \qquad R_R=[k_2], \qquad R_S=[k_3] \, .
	\end{align}
Consequently,
	\begin{align}
		R_Q^t=[1^{k_1}], \qquad R_R^t=[1^{k_2}], \qquad R_S^t=[1^{k_3}] \, .
	\end{align}
This is the representation-theoretic appearance of the determinant source. After the transpose, one-row representations become one-column, antisymmetric representations.
Therefore, the open-character factor now becomes simply
	\begin{align}
		 \chi_{[k_1]}(U^{-1}) \chi_{[k_2]}(V^{-1}) \chi_{[k_3]}(W^{-1}) \, = \, u^{-k_1}v^{-k_2}w^{-k_3} \, .
	\end{align}

For the one-column representation of the symmetric group $S_k$, the character is the sign representation:
	\begin{align}
		\chi_{[1^k]}(\pi)=\sgn(\pi) , \qquad \pi\in S_k \, .
	\end{align}
Thus, the coefficient in the one-giant sector becomes
	\begin{align}
		\mathcal{V}_{\mathcal{R};[1^{k_1}],[1^{k_2}],[1^{k_3}]} \, &= \, \frac{1}{k_1!k_2!k_3!} \sum_{\pi_1\in S_{k_1}} \sum_{\pi_2\in S_{k_2}} \sum_{\pi_3\in S_{k_3}} \sgn(\pi_1)\sgn(\pi_2)\sgn(\pi_3) 
		\sum_{\tau\in M_{k_1,k_2,k_3}} \chi_{\mathcal{R}}(\pi\tau) \, .
	\end{align}
Therefore, the $Q=R=S=1$ specialization of the character expansion is
	\begin{align}
		Z_{\mathrm V}^{(1,1,1)} \, &= \,  \sum_{k_{1},k_{2},k_{3}\ge0} \left(\frac{g}{2N}\right)^k \left( \frac{\kappa_1}{u} \right)^{k_1} \left( \frac{\kappa_2}{v} \right)^{k_2} \left( \frac{\kappa_3}{w} \right)^{k_3}
		\sum_{\substack{\mathcal{R}\vdash 2k\\ \ell(\mathcal{R})\le N}} \chi_{\mathcal{R}}(F) \, \mathcal{V}_{\mathcal{R};[1^{k_1}],[1^{k_2}],[1^{k_3}]}.
	\end{align}
This is already the character-expansion form of the single-giant determinant correlator.  The remaining work is to evaluate the finite-$N$ color factor.

Now we introduce the standard notation for the three channels:
	\begin{align}
		k_1=m+l \, , \qquad k_2=m+n \, , \qquad k_3=n+l.
	\end{align}
We also define the three finite-$N$ variables by the three effective propagator weights,
	\begin{align}
		x \, = \, \frac{g}{2}\frac{\kappa_1\kappa_2}{uv} \, , \qquad y \, = \, \frac{g}{2}\frac{\kappa_2\kappa_3}{vw} \, , \qquad \, z \, = \, \frac{g}{2}\frac{\kappa_3\kappa_1}{wu} \, .
	\end{align}
Since $k=m+n+l$, the full prefactor in the specialized character expansion satisfies
	\begin{align}
		\left(\frac{g}{2N}\right)^k \left( \frac{\kappa_1}{u} \right)^{k_1} \left( \frac{\kappa_2}{v} \right)^{k_2} \left( \frac{\kappa_3}{w} \right)^{k_3} \, = \, \frac{x^m y^n z^l}{N^{m+n+l}} \, .
	\end{align}
Thus the character expansion takes the form
	\begin{align}
		Z_{\mathrm V}^{(1,1,1)} \, = \, \sum_{m,n,l\ge0} \frac{\mathcal{I}_N(m,n,l)}{N^{m+n+l}} \, x^m y^n z^l \, ,
	\end{align}
where the remaining color factor is
	\begin{align}
		\mathcal{I}_N(m,n,l) \, = \,  \sum_{\substack{\mathcal{R}\vdash 2(m+n+l)\\ l(\mathcal{R})\le N}} \chi_{\mathcal{R}}(F) \, \mathcal{V}_{\mathcal{R};[1^{m+l}],[1^{m+n}],[1^{n+l}]} \, .
	\end{align}

The finite-$N$ polynomial is the unrefined color-counting result, so we set
	\begin{align}
		F \, = \, I_N \, .
	\end{align}
Then, Schur-Weyl duality gives the identity
	\begin{align}
		\Tr\left(\sigma I_N^{\otimes d}\right) \, = \, \sum_{\substack{\mathcal{R}\vdash d\\ \ell(\mathcal{R})\le N}} \chi_{\mathcal{R}}(I_N)\chi_{\mathcal{R}}(\sigma) \, = \, N^{C(\sigma)} \, ,
	\end{align}
where $C(\sigma)$ is the number of cycles of the permutation $\sigma$.  Substituting the antisymmetric coefficient into the closed-character sum gives
	\begin{align}
		\mathcal{I}_N(m,n,\ell) \, = \, \frac{1}{k_1!k_2!k_3!} \sum_{\pi_1\in S_{k_1}} \sum_{\pi_2\in S_{k_2}} \sum_{\pi_3\in S_{k_3}} \sgn(\pi_1)\sgn(\pi_2)\sgn(\pi_3)
		\sum_{\tau\in M_{k_1,k_2,k_3}} N^{C(\pi\tau)}.
	\end{align}
This formula is the antisymmetric color-loop sum.  It is still a character-expansion result, but the closed representation $\mathcal{R}$ has been summed explicitly using Schur-Weyl duality.

We now evaluate $\mathcal{I}_N(m,n,l)$ combinatorially. Since
\begin{align}
    N^{C(\pi\tau)}=\Tr (\pi\t I^{\otimes d}_{N})\,,
\end{align}
we have
\begin{align}
		\mathcal{I}_N(m,n,\ell) \, = \, \frac{1}{k_1!k_2!k_3!} \sum_{\pi_1\in S_{k_1}} \sum_{\pi_2\in S_{k_2}} \sum_{\pi_3\in S_{k_3}}\sum_{\tau\in M_{k_1,k_2,k_3}} \Tr[\sgn(\pi_1)\pi_{1}\sgn(\pi_2)\pi_{2}\sgn(\pi_3)\pi_{3}\tau I_{N}^{\otimes d}]\,,
	\end{align}
 where we defined $\pi_{\a}$ for $\a=1,2,3$ by acting on the block $B_{\a}$ and trivially on the other blocks.
 This is equivalent to considering the Wick contractions of principal minors $e_{k_{\a}}(I_{N})$. We can analyze this similarly to section~\ref{sec:v3}. 
Combining the label count, endpoint bijections, and sign, we obtain
	\begin{align}
		\mathcal{I}_N(m,n,l) \, = \, \sum_{\substack{0\le t\le\min(m,n,l)\\ m+n+l-t\le N}} \frac{(-1)^tN!\,m!\,n!\,l!}{(N-m-n-l+t)! (m-t)!(n-t)!(l-t)!(t!)^2}.
	\end{align}
Since the character expansion had the overall factor $N^{-(m+n+l)}$, the coefficient of $x^m y^n z^l$ is
	\begin{align}
		C_N(m,n,l) \, = \, \frac{\mathcal{I}_N(m,n,l)}{N^{m+n+l}} \, .
	\end{align}
Therefore, we find
	\begin{align}
		Z_N(x,y,z) \, = \, \sum_{m,n,l\ge0} C_N(m,n,l) x^m y^n z^l \, ,
	\end{align}
with
	\begin{align}
		C_N(m,n,l) \, = \, \sum_{\substack{0\le t\le\min(m,n,l)\\ m+n+l-t\le N}} (-1)^t \frac{N!\,m!\,n!\,l!}{(N-m-n-l+t)! (m-t)!(n-t)!(l-t)!(t!)^2 N^{m+n+l}} \, .
	\end{align}
This is the exact single-giant finite-$N$ polynomial for the V-type three-point function obtained in section~\ref{sec:v3}.

\section{String Construction}
\label{sec:string}
In this section, we discuss a conservative extension of the Gopakumar-Kaushik-Komatsu-Mazenc-Sarkar (GKKMS) \cite{Gopakumar:2024jfq} string construction
from the two-matrix model to the three-matrix model used in the previous sections. 
The main conclusion is that the worldsheet part of the construction generalizes directly.
Each connected Wick contraction of the three-matrix V-type model determines a three-colored ribbon graph, hence an integer Strebel point in closed-string moduli space. 
The target-space part is subtler.
The ordinary one-propagator/one-sheet Belyi map of the bipartite two-matrix model should be replaced either by a barycentrically subdivided Belyi map, whose degree is twice the number of propagators,
or by a colored Hurwitz/constellation description retaining the three insertion colors as separate branch data.

The starting point is the three-matrix V-type model introduced in section~\ref{sec:triality} whose action is 
	\begin{align}
		S[M_1,M_2,M_3] \, = \, \frac{1}{2}\sum_{\a,\b=1}^3 K_{\a\b} \Tr(\sqrt{Y}M_\a \sqrt{Y}M_\b) \, ,
	\end{align}
where the color-space kernel is the $3\times 3$ matrix
	\begin{align}
		K_{\alpha\beta} \, = \, - \frac{\delta_{\alpha\beta}}{\kappa_\alpha^2} + \frac{1-\delta_{\alpha\beta}}{\kappa_\alpha\kappa_\beta} \, .
	\end{align}
Then, the propagator is given by
	\begin{align}
		(K^{-1})_{\a\a} \, = \, 0 \, , \qquad (K^{-1})_{\a\b} \, = \, \frac{\kappa_\a \kappa_\b}{2} \quad (\a \ne \b) \, .
	\end{align}
Thus the Wick contractions have the form
	\begin{align}
		\big\langle (M_\a)_{ij}(M_\b)_{kl}\big\rangle \, = \, \frac{G_{\a\b}}{N}\, \frac{\delta_{il}\delta_{jk}}{\sqrt{y_i y_j}} \, , \qquad
		G_{\a\a} \, = \, 0 \, , \qquad G_{\a\b} \, = \, \frac{g}{2}\kappa_\a \kappa_\b \, ,
	\end{align}
where we assume $\a\neq\b$ in the third relation.
If $Y={\bf 1}$ and the $\kappa_\a$ are absorbed into the matrices, this reduces to the simplified V-type propagator $G_{\a\b}=g/2$ for $\a\ne\b$.
The determinant insertions or, equivalently, their Miwa expansion, generate trace vertices of the form
	\begin{align}
		&\prod_{a=1}^{Q}\det(u_a-M_1) \prod_{b'=1}^{R}\det(v_b'-M_2) \prod_{\hat{c}=1}^{S}\det(w_{\hat{c}}-M_3) \nn\\
		&\quad=\exp\left[-\sum_{m\geq 1}\frac{t_{1m}}{m}\Tr M_1^m -\sum_{m\geq 1}\frac{t_{2m}}{m}\Tr M_2^m -\sum_{m\geq 1}\frac{t_{3m}}{m}\Tr M_3^m\right].
	\end{align}
up to the usual source determinants.  Therefore, a Feynman diagram has vertices of three colors, labelled $1,2,3$, and edges only between different colors.  It is a three-colored, or tripartite, ribbon graph.  It need not be bipartite: triangular cycles of type $1\to2\to3\to1$ are allowed.

Let $\Gamma$ be a connected Wick-contraction ribbon graph. 
We also denote by $V_\a$ the number of trace vertices of color $\a$, and by $V=V_1+V_2+V_3$ the total number of trace vertices. 
In addition, let $E_{\a\b}$ be the number of propagators of type $\a\b$, and let $E = E_{12} + E_{23} + E_{13}$.
If the total number of matrix letters at the three colors is $k_1,k_2,k_3$, then Wick contractions require
	\begin{align}
		k_1 \, = \, E_{12}+E_{13} \, , \qquad k_2 \, = \, E_{12}+E_{23} \, , \qquad k_3 \, = \, E_{23}+E_{13} \, .
	\end{align}
Equivalently,
	\begin{align}
		E_{12} \, = \, \frac{k_1+k_2-k_3}{2} \, , \qquad E_{23} \, = \, \frac{k_2+k_3-k_1}{2} \, , \qquad E_{13} \, = \, \frac{k_3+k_1-k_2}{2} \, .
	\end{align}
Hence the three triangle inequalities must hold and $k_1+k_2+k_3$ must be even.  These are exactly the three-color analogues of the bipartite length constraints in GKKMS with the two-matrix model.
The power of $N$ associated with an unnormalized connected diagram is
	\begin{align}
		N^{F-E} \, = \, N^{2-2h-V} \, ,
	\end{align}
where $F$ is the number of index loops/faces of the ribbon graph and $V-E+F = 2-2h$.
Thus the same genus expansion as in GKKMS survives unchanged, with $g_s\sim 1/N$. 
The only difference is that the edges carry a type label $\alpha\beta$ and therefore a type-dependent weight:
	\begin{align}
		W(\Gamma) \, \propto \, \frac{1}{|\textrm{Aut}\Gamma|} \, N^{-E} \prod_{\alpha<\beta}G_{\alpha\beta}^{E_{\alpha\beta}} \, \prod_{f}\Tr Y^{-m_f/2} \, ,
	\end{align}
where $m_f$ is the number of propagator sides bordering the face $f$. 
In the two-matrix model all faces have even length, but in the three-matrix model odd faces can occur. 
Thus, it is useful to use the notation
	\begin{align}
		\widetilde{s}_m \, := \, \Tr Y^{-m/2}
	\end{align}
for face weights.  This is the same point that appears in the F-type formula through powers such as $s_{m/2}$.

\subsection{Worldsheet construction}
The worldsheet part of the GKKMS construction uses only a metrized ribbon graph. 
Since every three-matrix Wick contraction still gives an oriented ribbon graph, the construction generalizes almost verbatim.  
In the following, we will make this statement a little bit more precise.

We start with a connected three-colored ribbon graph $\G$. 
The cyclic ordering at each trace vertex is inherited from the order of matrix letters in the trace, and the double-line propagators determine the faces. 
Assign unit length to every propagator. If desired, first pass to the skeleton graph $\G_\textrm{sk}$ by bundling homotopic parallel propagators. 
A skeleton edge $\bar{e}$ has length
	\begin{align}
		\ell_{\bar{e}} \, = \, \#\{\hbox{unit propagators in the bundle }\bar{e}\} \in \mathbb{Z}_{>0} \, .
	\end{align}
Since the original trace vertices are to become closed-string vertex insertions, one should not directly identify $\G_\textrm{sk}$ with the Strebel graph. 
Rather, as in GKKMS, one identifies the dual ribbon graph with the Strebel graph:
	\begin{align}
		\mathcal{S}_\G \, := \, \G_\textrm{sk}^{\vee} \, ,
	\end{align}
where the symbol $\vee$ means the dual ribbon graph.
Faces of $\mathcal{S}_\G$ are in one-to-one correspondence with the original trace vertices of $\G$. 
Thus every trace insertion becomes a marked point on the closed worldsheet. 
The color of the original trace vertex becomes a label attached to the corresponding marked point. 
If $p_a$ is the perimeter of the face of $\mathcal{S}_\G$ associated with the trace vertex $a$, then
	\begin{align}
		p_a=\sum_{\bar e\subset \partial f_a}\ell_{\bar e} \, = \, \hbox{number of matrix letters in the trace at }a.
	\end{align}
The Strebel theorem then gives a unique point of decorated moduli space,
	\begin{align}
		(\mathcal{S}_\G,\{\ell_{\bar e}\}) \quad \longleftrightarrow\quad (\Sigma_{h,V};p_1,\ldots,p_V;\; \hbox{perimeters }p_a),
	\end{align}
where $\Sigma_{h,V}$ is a genus-$h$ Riemann surface with $V$ marked points. 
This is the key sense in which the three-matrix Feynman diagrams do not discretize the worldsheet itself.  They label integer points in the Strebel cell decomposition of moduli space.

The same construction has a more physical open-string interpretation. Each propagator is a unit strip. 
A strip of type $\a\b$ has one end attached to an $M_\a$ trace vertex and the other to an $M_\b$ trace vertex. 
The two long boundaries of the strip are pieces of index loops. 
Gluing the strip ends according to the cyclic order at the trace vertices and gluing the long boundaries according to the index loops produces a closed oriented surface.

In the two-matrix GKKMS model, the graph is bipartite, so the strip ends alternate between the two colors. 
In the three-matrix model, this alternating rule should be replaced by the following more general rule:
	\begin{align}
		\hbox{at an }M_\a & \hbox{ trace vertex, glue cyclically all strip ends of types }\a\b\hbox{ and }\a\g \, ,
	\end{align}
where $\{\a,\b,\g\}=\{1,2,3\}$.
No same-color strip is allowed, but the cyclic sequence around an $M_\a$ trace can mix the two possible neighboring colors. 
This is why triangular cycles and odd index-loop faces are allowed. 
Nothing in the Strebel reconstruction forbids such graphs: trivalent and higher-valent Strebel vertices are standard,
and odd-valent vertices simply correspond to the appropriate order of zero of the Strebel differential.
Schematically, the three-matrix worldsheet map is therefore
	\begin{align}
	\begin{gathered}
		\hbox{3-colored V-type Wick graph }\Gamma \quad \Longrightarrow \quad \hbox{dual skeleton }\mathcal{S}_\G \\[2pt]
		\Longrightarrow \quad \hbox{integer Strebel surface }(\Sigma_{h,V},\phi_S)
	\end{gathered}
	\end{align}
    with the Strebel differential $\phi_{S}$.
The marked points carry color labels $1,2,3$, and the Strebel edges may carry the type label of the propagator bundle that they cross.

If the target-space or worldsheet action assigns an area $A_{\a\b}$ to a unit strip of type $\a\b$, then the action of a diagram should be of the form
	\begin{align}
		S_{\rm ws}(\G) \, = \, \sum_{\a<\b}E_{\a\b}A_{\a\b} \, .
	\end{align}
Matching the matrix-model propagator weights requires
	\begin{align}
		e^{-A_{\a\b}} \, \propto \, G_{\a\b} \, ,
	\end{align}
with the usual understanding that signs and phases may be absorbed into contour choices or topological-string weights. 
When all off-diagonal propagators are equal, the worldsheet action is simply proportional to the total number of strips $E$. 
When the $\k_\a$ or spacetime three-point data are kept, the string action must remember the edge type.
The $Y$-source has the same interpretation as in GKKMS: it weights index-loop faces, or equivalently vertices of the dual Strebel graph. 
The only new feature is that the non-bipartite graph can have odd face length, hence the natural variables are $\widetilde{s}_m=\Tr Y^{-m/2}$ rather than only $s_k=\Tr Y^{-k}$.

The F-type three-point integral introduced in section~\ref{sec:triality} has three bifundamental bosonic matrices. 
Its interaction is governed by a block off-diagonal matrix $\L$ whose trace powers have the schematic form
	\begin{align}
		\Tr\Lambda^m \, = \, \sum_{\substack{i_1,\ldots,i_m\in\{1,2,3\}\\ i_{r+1}\ne i_r,\; i_{m+1}=i_1}} \Tr\!\big(X_{i_1 i_2}X_{i_2 i_3}\cdots X_{i_m i_1}\big).
	\end{align}
Thus, the F-type model sums closed walks on a three-node quiver with no self-step. 
This is the open-string counterpart of the V-type statement $G_{\a\a}=0$.
From the closed-string point of view, the V-type and F-type descriptions should be interpreted as two open-string cell decompositions of the same closed Strebel surfaces. 
In the two-matrix GKKMS case, this is the open-closed-open triality:
the two open descriptions are related by graph duality or partial graph duality, while the closed description is the common Strebel worldsheet. 
The three-point character comparison in section~\ref{sec:character} supports the analogous statement
	\begin{align}
	\begin{gathered}
		\hbox{3-colored V-type open graphs} \quad\Longleftrightarrow\quad \hbox{closed integer-Strebel worldsheets} \\[2pt]
		\Longleftrightarrow \quad \hbox{3-node F-type quiver open graphs}.
	\end{gathered}
	\end{align}
This statement is strongest at the level of the combinatorial expansion and the resulting closed-surface data. 
A fully local target-space sigma-model derivation would require specifying the analogue of the GKKMS target metric and map for the three-color problem.

\subsection{Target space construction}
The GKKMS target-space construction is more special than the worldsheet construction. 
In the two-matrix model, bipartiteness produces a dessin d'enfant~\cite{Grothendieck:1984edp,Lando:2004gsa}: there are two types of vertices and one type of face, leading naturally to a Belyi map~\cite{MR534593,Lando:2004gsa} with three branch values. 
For the three-matrix model, there are instead three vertex colors plus faces.  Therefore, the literal GKKMS Belyi construction does not carry over with one sheet per propagator.

There are two natural replacements we can consider.

First, one can forget the three colors temporarily and use the standard barycentric Belyi map associated with any ribbon graph. 
Let $H$ be the set of half-edges, thus
	\begin{align}
		|H| \, = \, d \, = \, 2E \, .
	\end{align}
Let $\s_V$ be the cyclic-order permutation around all original vertices, let $\t$ be the fixed-point-free involution pairing half-edges into propagators, and let $\s_f$ be the face permutation. 
Then,
	\begin{align}
		\s_f\, \t \, \s_V \, = \, 1 \, .
	\end{align}
This gives an ordinary Belyi map of degree $2E$, not degree $E$. The three colors survive as labels on the cycles of $\s_V$.

Second, to keep the three insertion points distinct in target space, one can use a colored Hurwitz or constellation description.\footnote{See~\cite{deMelloKoch:2010hav} for the example of the analysis using a colored Hurwitz description.} 
Let $\s_\a$ be the cyclic-order permutation around vertices of color $\a$, extended by fixed points on all half-edges not incident on vertices of color $\a$.  Then, the natural relation is
	\begin{align}
		\s_f \, \t \, \s_3 \s_2 \s_1 \, = \, 1 \, .
	\end{align}
This is a cover with branch data associated with the three colors, the edge pairing, and the faces. 
Since fixed points are included, the Riemann-Hurwitz formula reproduces the ribbon-graph Euler characteristic:
	\begin{align}
		2-2h \, &= \, 2d - \left[ \sum_{\a=1}^3 \big( d-C(\s_\a) \big) + \big( d-C(\t) \big) + \big( d-C(\s_f) \big) \right] \nn\\
		&= V-E+F.
	\end{align}
Here $C(\pi)$ denotes the number of cycles of a permutation $\pi$, and $C(\t)=E$.  The equality also uses
	\begin{align}
		C(\s_1) + C(\s_2) + C(\s_3) \, = \, V+2d \, ,
	\end{align}
because each half-edge is acted on by exactly one of the three vertex-color permutations and is fixed by the other two.

This gives the right generalization of the GKKMS string construction: the localization to integer Strebel points survives, but the target-space map is naturally a colored constellation, or a barycentric Belyi map with additional color decoration, rather than the original two-color dessin with one sheet per propagator.

\section{Conclusions and Discussions}
\label{sec:conclusions}
We have developed a three-point generalization of open--closed--open triality for protected half-BPS correlators in four-dimensional $\mathcal{N}=4$ SYM. 
Starting from determinant operators representing giant gravitons, the V-type description is a Gaussian three-matrix model whose trace vertices carry three colors. 
By rewriting the determinants with auxiliary fermions, integrating out the color matrices, and performing Hubbard--Stratonovich transformations, we obtained an F-type theory of three bifundamental matrices on a triangular quiver. 
Its interactions are naturally organized by closed quiver walks.  In contrast to the bipartite two-point model, this quiver admits oppositely oriented cubic cycles, which encode genuinely three-point index flow.

We tested the triality at several complementary levels. 
In the single-giant sector, the V-type model yields exact finite-$N$ polynomials for both two- and three-point functions. 
For the sourceless three-point function, the tree-level and one-loop contributions around the F-type saddle manifolds reproduce the leading large-$N$ asymptotics of the exact even-$N$ result. 
We also derived character expansions of both open descriptions. 
An explicit bijection between V-type Wick-contraction data and F-type closed-walk data identifies their representation coefficients,
while the restriction to one determinant at each point independently reproduces the exact single-giant polynomial.

At the closed-string corner, every connected V-type contraction defines a three-colored ribbon graph. 
After skeletonization and dualization, it determines an integer Strebel surface and hence a discrete point in decorated moduli space. 
The worldsheet construction therefore extends directly beyond the bipartite case. 
The target-space interpretation is less rigid: the original one-propagator/one-sheet Belyi description must be replaced either by a barycentrically subdivided Belyi map or by a colored Hurwitz constellation. 
At present, this is a combinatorial proposal rather than a derivation of a local target-space sigma model.

Several directions deserve further study. 
First, one should try to solve the multi-giant F-type model with general $(Q,R,S)$, including its saddle structure, fluctuation determinants, and finite-$N$ character sums. 
Second, deriving loop equations and a spectral curve would test whether the full genus expansion is governed by topological recursion \cite{Eynard:2007kz}. 
Third, replacing determinant ground states by operator bases for giant gravitons with attached open strings would extend the triality beyond the protected configurations considered here \cite{deMelloKoch:2007rqf,deMelloKoch:2007nbd}.  
Finally, since the supersymmetric circular Maldacena-Wilson loop operators \cite{Maldacena:1998im} are half-BPS and they are reduced to a Gaussian matrix model \cite{Pestun:2007rz}, one can try to formulate a Wilson-loop open–closed–open triality. 
These developments may turn the present combinatorial correspondence into a more complete worldsheet and target-space formulation of open--closed--open triality.

\section*{Acknowledgements}

We are grateful to Edward Mazenc, Yuji Okawa and Tadashi Takayanagi for useful discussion.
The work of KK is supported in part by JSPS KAKENHI Grant Number JP25KJ0997.
The work of KS is supported by JSPS KAKENHI Grant No.~23K13105. 
We acknowledge OpenAI’s ChatGPT for assistance with calculations, manuscript editing, and literature searches.
All AI-assisted calculations, statements, and references were independently reviewed and verified by the authors, who take full responsibility for the content of the manuscript.

\appendix
\section{Detailed Derivation of F-type model for three-point function}
In this appendix, we present the detailed calculations that appear in section~\ref{sec:threept-MatrixDuality}.
\subsection{Integrating out bosons}\label{sec:completing-square}
Let us begin with~\eqref{eq:integrating-in-fermion}:
\begin{align}
        Z_{\mathrm{3det}}&=\mathcal{K}\int dM_{1}dM_{2}dM_{3}d\eta d\eta^{\dagger} d\psi d\psi^{\dagger} d\chi d\chi^{\dagger}\exp\bigg[-\frac{N}{g}S+\eta^{\dagger}_{ia}(U_{ab}\delta_{ij}-\delta_{ab}M_{1,ij})\eta_{jb} \nn\\
        &\qquad+\psi^{\dagger}_{ia'}(V_{a'b'}\delta_{ij}-\delta_{a'b'}M_{2,ij})\psi_{jb'}+\chi^{\dagger}_{i\hat{a}}(W_{\hat{a}\hat{b}}\delta_{ij}-\delta_{\hat{a}\hat{b}}M_{3,ij})\chi_{j\hat{b}}\bigg]
\end{align}
This can be deformed into
\begin{align}
        Z_{\mathrm{3det}}
        &=\mathcal{K}\int dM_{1}dM_{2}dM_{3}d\eta d\eta^{\dagger} d\psi d\psi^{\dagger} d\chi d\chi^{\dagger}\exp\Big(\eta^{\dagger}_{ia}U_{ab}\eta_{ib}+\psi^{\dagger}_{ia'}V_{a'b'}\psi_{ib'}+\chi^{\dagger}_{i\hat{a}}W_{\hat{a}\hat{b}}\chi_{i\hat{b}} +I\Big)\,,
\end{align}
with
\begin{align}
    I=-\frac{N}{2g}\Tr\left(\sqrt{Y}M_{\alpha}K_{\alpha\beta}\sqrt{Y}M_{\beta}\right)-\eta^{\dagger}_{ia}M_{1,ij}\eta_{ja}-\psi^{\dagger}_{ia'}M_{2,ij}\psi_{ja'}-\chi^{\dagger}_{i\hat{a}}M_{3,ij}\chi_{j\hat{a}}
\end{align}
For simplicity, we introduce
\begin{align}
    \widetilde{M}_{\alpha}=Y^{\frac{1}{4}}M_{\alpha}Y^{\frac{1}{4}}\,,
\end{align} 
and
\begin{align}
    (B_{1})_{ij}=Y^{-\frac{1}{4}}_{ik}\eta_{ka}\eta^{\dagger}_{la}Y^{-\frac{1}{4}}_{lj}\,,\qquad (B_{2})_{ij}=Y^{-\frac{1}{4}}_{ik}\psi_{ka'}\psi^{\dagger}_{la'}Y^{-\frac{1}{4}}_{lj}\,,\qquad (B_{3})_{ij}=Y^{-\frac{1}{4}}_{ik}\chi_{k\hat{a}}\chi^{\dagger}_{l\hat{a}}Y^{-\frac{1}{4}}_{lj}\,.
\end{align}
Then, the term $I$ can be simplified as
\begin{align}
    I=-\frac{N}{2g}\Tr\,(\widetilde{M}^{T}K\widetilde{M})+\Tr\,(B^{T}\widetilde{M})\,,
\end{align}
where
\begin{align}
    \widetilde{M}^{T}=(\widetilde{M}_{1},\widetilde{M}_{2},\widetilde{M}_{3})\,,\qquad B^{T}=(B_{1},B_{2},B_{3})\,,
\end{align}
and the trace is taken over the indices $\alpha=1,2,3$ and $i=1,\ldots,N$.
By completing the square, we can then carry out the Gaussian integral:
\begin{align}
    \begin{split}
        Z_{\mathrm{3det}}
        &=\frac{1}{(\det_{Q}U)^{N}(\det_{R}V)^{N}(\det_{S}W)^{N}}\int d\eta d\eta^{\dagger} d\psi d\psi^{\dagger} d\chi d\chi^{\dagger} \\
        &\qquad \times \exp\Big(\eta^{\dagger}_{ia}U_{ab}\eta_{ib}+\psi^{\dagger}_{ia'}V_{a'b'}\psi_{ib'}
        + \chi^{\dagger}_{i\hat{a}}W_{\hat{a}\hat{b}}\chi_{i\hat{b}}+\frac{g}{2N}\Tr (B_{\alpha}K^{-1}_{\alpha\beta}B_{\beta})\Big)
    \end{split}
\end{align}
with
\begin{align}
    K^{-1}=\frac{1}{2}\begin{pmatrix}
        0 & \kappa_{1}\kappa_{2} & \kappa_{1}\kappa_{3}\\
        \kappa_{2}\kappa_{1} & 0 & \kappa_{2}\kappa_{3}\\
        \kappa_{3}\kappa_{1}& \kappa_{3}\kappa_{2} & 0
      \end{pmatrix}
      \,.
\end{align}

\subsection{Integrating in Bosons}\label{sec:integrate-in}
Let us next integrate in the bosons to derive~\eqref{eq:interating-in-boson}.
We introduce
\begin{equation}
    \begin{split}
    &\lambda^{1}_{a'a}=\psi^{\dagger}_{la'}Y^{-\frac{1}{2}}_{li}\eta_{ia}\,,\qquad \lambda^{2}_{\hat{a}a'}=\chi^{\dagger}_{l\hat{a}}Y^{-\frac{1}{2}}_{li}\psi_{ia'}\,,\qquad \lambda^{3}_{a\hat{a}}=\eta^{\dagger}_{la}Y^{-\frac{1}{2}}_{li}\chi_{i\hat{a}}\,,\\
    &\widetilde\lambda^{1}_{aa'}=\eta^{\dagger}_{la}Y^{-\frac{1}{2}}_{li}\psi_{ia'}\,,\qquad \widetilde\lambda^{2}_{a'\hat{a}}=\psi^{\dagger}_{la'}Y^{-\frac{1}{2}}_{li}\chi_{i\hat{a}}\,,\qquad \widetilde\lambda^{3}_{\hat{a}a}=\chi^{\dagger}_{l\hat{a}}Y^{-\frac{1}{2}}_{li}\eta_{ia}\,.
    \end{split}
\end{equation}
We note that $\widetilde\lambda^{1}$ is $Q\times R$, $\widetilde\lambda^{2}$ is $R\times S$, and $\widetilde\lambda^{3}$ is $S\times Q$ matrices. By considering Gaussian integrals, we have
\begin{equation}
    \begin{split}
        e^{\Tr\frac{g}{2N}B_{\a}K_{\a\b}^{-1}B_{\b}}
        &=e^{-\frac{g}{2N}\Tr(\kappa_{1}\kappa_{2}\widetilde\lambda_{1}\lambda_{1}+\kappa_{2}\kappa_{3}\widetilde\lambda_{2}\lambda_{2}+\kappa_{3}\kappa_{1}\widetilde\lambda_{3}\lambda_{3})}\\
        &=\left(\frac{N}{2\pi g}\right)^{QR+RS+SQ}\int \prod_{\alpha=1}^{3} d{\Sigma}_{\alpha}d{\Sigma}_{\alpha}^{\dagger}\\
        &\times\exp\Big\{-\frac{1}{2}\Big[\Tr\left(\frac{N}{g}{\Sigma}_{1\,a'a}^{\dagger}{\Sigma}_{1\,aa'}-i\sqrt{\kappa_{1}\kappa_{2}}\widetilde\lambda^{1}_{aa'}{\Sigma}_{1\,aa'}-i\sqrt{\kappa_{1}\kappa_{2}}{\Sigma}_{1\,a'a}^{\dagger}\lambda_{a'a}^{1}\right)\\
        &+\Tr\left(\frac{N}{g}{\Sigma}_{2\,\hat{a}a'}^{\dagger}{\Sigma}_{2\,a'\hat{a}}-i\sqrt{\kappa_{2}\kappa_{3}}\widetilde\lambda^{2}_{a'\hat{a}}{\Sigma}_{2\,a'\hat{a}}-i\sqrt{\kappa_{2}\kappa_{3}}{\Sigma}_{2\,\hat{a}a'}^{\dagger}\lambda_{\hat{a}a'}^{2}\right) \\
        &+\Tr\left(\frac{N}{g}{\Sigma}_{3\,a\hat{a}}^{\dagger}{\Sigma}_{3\,\hat{a}a}-i\sqrt{\kappa_{3}\kappa_{1}}\widetilde\lambda^{3}_{\hat{a}a}{\Sigma}_{3\,\hat{a}a}-i\sqrt{\kappa_{3}\kappa_{1}}{\Sigma}_{3\,a\hat{a}}^{\dagger}\lambda_{a\hat{a}}^{3}\right)\Big]\Big\}\,.
    \end{split}
\end{equation}
We then have
\begin{align}
    Z_{\mathrm{3det}}=\mathcal{K}_{3}\int \prod_{i=1}^{3} d\Sigma_{i}d\Sigma_{i}^{\dagger}\,d\eta d\eta^{\dagger} d\psi d\psi^{\dagger} d\chi d\chi^{\dagger}\exp\left(\widetilde{I}[\Sigma_{i},\Sigma_{i}^{\dagger},\eta,\psi,\chi]\right)\,,
\end{align}
with
\begin{align}
    \mathcal{K}_{3}=\frac{1}{(\det_{Q}U)^{N}(\det_{R}V)^{N}(\det_{S}W)^{N}}\left(\frac{N}{2\pi g}\right)^{QR+RS+SQ}\,,
\end{align}
\begin{equation}
    \begin{split}
    \widetilde{I}[\Sigma_{i},\Sigma_{i}^{\dagger},\eta,\psi,\chi]&=-\frac{N}{2g}\left[\Tr(\Sigma_{1\,a'a}^{\dagger}\Sigma_{1\,aa'}+\Sigma_{2\,\hat{a}a'}^{\dagger}\Sigma_{2\,a'\hat{a}}+\Sigma_{3\,a\hat{a}}^{\dagger}\Sigma_{3\,\hat{a}a})\right]\\
    &\qquad+\sum_{l}
    \begin{pmatrix}
        \eta_{la}^{\dagger} & \psi_{la'}^{\dagger} & \chi_{l\hat{a}}^{\dagger} 
    \end{pmatrix}
    \,A_{l\{a,a',\hat{a}\}\{b,b',\hat{b}\}}\,
    \begin{pmatrix}
        \eta_{lb} \\ \psi_{lb'} \\ \chi_{l\hat{b}} 
    \end{pmatrix}\,,
    \end{split}
\end{equation}
and 
\begin{align}
    A_{l}=
    \begin{pmatrix}
        U_{ab} & \frac{i}{2}\sqrt{\kappa_{1}\kappa_{2}}y^{-\frac{1}{2}}_{l}\Sigma_{1\,ab'} & \frac{i}{2}\sqrt{\kappa_{1}\kappa_{3}}y^{-\frac{1}{2}}_{l}\Sigma^{\dagger}_{3\,a\hat{b}} \\
        \frac{i}{2}\sqrt{\kappa_{1}\kappa_{2}}y^{-\frac{1}{2}}_{l}\Sigma^{\dagger}_{1\,a'b} & V_{a'b'} &  \frac{i}{2}\sqrt{\kappa_{2}\kappa_{3}}y^{-\frac{1}{2}}_{l}\Sigma_{2\,a'\hat{b}}\\
        \frac{i}{2}\sqrt{\kappa_{1}\kappa_{3}}y^{-\frac{1}{2}}_{l}\Sigma_{3\,\hat{a}b} & \frac{i}{2}\sqrt{\kappa_{2}\kappa_{3}}y^{-\frac{1}{2}}_{l}\Sigma^{\dagger}_{2\,\hat{a}b'}& W_{\hat{a}\hat{b}}
    \end{pmatrix}\,.
\end{align}
This is the desired result~\eqref{eq:interating-in-boson}. The matrix $A_{l}$ can be deformed into
\begin{equation}
    \begin{split}
        A_{l}=
        \begin{pmatrix}
        U & 0 & 0 \\
        0 & V & 0\\
        0 & 0 & W
    \end{pmatrix}
    \begin{pmatrix}
        I_{Q\times Q} & \frac{i}{2}\sqrt{\kappa_{1}\kappa_{2}}y^{-\frac{1}{2}}_{l}U^{-1}\Sigma_{1} & \frac{i}{2}\sqrt{\kappa_{1}\kappa_{3}}y^{-\frac{1}{2}}_{l}U^{-1}\Sigma^{\dagger}_{3} \\
        \frac{i}{2}\sqrt{\kappa_{1}\kappa_{2}}y^{-\frac{1}{2}}_{l}V^{-1}\Sigma^{\dagger}_{1} & I_{R\times R} &  \frac{i}{2}\sqrt{\kappa_{2}\kappa_{3}}y^{-\frac{1}{2}}_{l}V^{-1}\Sigma_{2}\\
        \frac{i}{2}\sqrt{\kappa_{1}\kappa_{3}}W^{-1}y^{-\frac{1}{2}}_{l}\Sigma_{3} & \frac{i}{2}\sqrt{\kappa_{2}\kappa_{3}}y^{-\frac{1}{2}}_{l}W^{-1}\Sigma^{\dagger}_{2}& I_{S\times S}
    \end{pmatrix}\,,
    \end{split}
\end{equation}
where $I_{Q\times Q}$ is a $Q$-dimensional identity matrix. We then have
\begin{align}
    A_{l}=\mathrm{diag}(U,V,W)\left(I_{(Q+R+S)\times(Q+R+S)}+\frac{i}{2}\,y_{l}^{-\frac{1}{2}}\,\Lambda\right)
\end{align}
with
\begin{align}
   \Lambda=
    \begin{pmatrix}
        0 & \sqrt{\kappa_{1}\kappa_{2}}U^{-1}\Sigma_{1} & \sqrt{\kappa_{1}\kappa_{3}}U^{-1}\Sigma^{\dagger}_{3} \\
        \sqrt{\kappa_{1}\kappa_{2}}V^{-1}\Sigma^{\dagger}_{1} & 0 & \sqrt{\kappa_{2}\kappa_{3}}V^{-1}\Sigma_{2}\\
        \sqrt{\kappa_{1}\kappa_{3}}W^{-1}\Sigma_{3} & \sqrt{\kappa_{2}\kappa_{3}}W^{-1}\Sigma_{2}^{\dagger} & 0
    \end{pmatrix}\,. 
\end{align}
Thus, the determinant simplifies as follows:
\begin{equation}
    \begin{split}
    \prod_{l}\mathrm{det}_{Q+R+S}\,A_{l}
    &=(\mathrm{det}_{Q}U\,\mathrm{det}_{R}V\,\mathrm{det}_{S}W)^{N}\,\prod_{l}\exp\Tr_{Q+R+S}\,\mathrm{log}\left(I+\frac{i}{2}\,y_{l}^{-\frac{1}{2}}\Lambda\right)\\
    &=(\mathrm{det}_{Q}U\,\mathrm{det}_{R}V\,\mathrm{det}_{S}W)^{N}\,\exp\left[-\sum_{k=1}^{\infty}\frac{1}{k}\left(-\frac{i}{2}\right)^{k}s_{\frac{k}{2}}\,\Tr\Lambda^{k}\right]\,.
    \end{split}
\end{equation}
We can now proceed the discussion discussed in section~\ref{sec:threept-MatrixDuality}.

\section{Technical Details of V-type Character Expansion}\label{sec:details-character}
In this section, we present the technical details of the V-type character expansions.

Let us first explain the derivation of~\eqref{eq:stab}. Since the action $\tau$ represents the Wick contraction, it is decomposed by the products of the transposition
\begin{align}
    \tau=(a_{1}a_{2})(a_{3}a_{4})\ldots(a_{d-1}a_{d})\,,
\end{align}
where $a_{i}$ with $i=1,2,\ldots,d$ are indices which are contracted, and if $a_{i}$ for odd $i$ is included in $B_{\a}$, $a_{i+1}$ is not included in $B_{\a}$. The action $\tau_{0}$ can be decomposed similarly. If the action $\tau_{0}$ contains the transposition $(p\,q)$ in the decomposition, we can verify
\begin{align}
    h(p\,q)h^{-1}=(h(p)\,h(q))\,.
\end{align}
This implies $h\tau_{0}h^{-1}\in M_{k_{1}k_{2}k_{3}}$. Moreover, since the action $h$ maps $B_{\a}$ to $B_{\a}$, $h\tau_{0}h^{-1}$ can represent all the elements of $M_{k_{1}k_{2}k_{3}}$. We then take the quotient by $\mathrm{Stab}_{H}(\tau_{0})$ to obtain~\eqref{eq:stab}.

Let us next consider~\eqref{eq:Schur-decomp}.
 We take $\Gamma_{R}(X_{k_{1},k_{2},k_{3}})=T_{rs}$, which is a map from $W_{r}\otimes M_{r}$ to $W_{s}\otimes M_{s}$. From the commutativity~\eqref{eq:commutativity}, we have
 \begin{align}
     T_{sr}\Gamma_{r}(h)=\Gamma_{s}(h)T_{sr}\,.
 \end{align}
 When we write
 \begin{align}
     T_{sr}=\sum_{a,b}A_{ab}\otimes X_{ab}\,,
 \end{align}
 where $X_{ab}$ maps the basis vectors of $M_{r}$ to those of $M_{s}$,
 we find
 \begin{align}
     A_{ab}\Gamma_{r}(h)=\Gamma_{s}(h)A_{ab}\,.
 \end{align}
 Since the representation $W_{\lambda}$ is irreducible,  we obtain
 \begin{align}
     A_{ab}=0
 \end{align}
 for $r\neq s$ and
 \begin{align}
     A_{ab}=c_{ab}I_{W_{r}}\,.
 \end{align}
 for $r=s$ with a constant $c_{ab}$ from Schur's lemma. We then have
 \begin{align}
     T_{rr}=I_{W_{r}}\otimes \sum_{a,b}c_{ab}X_{ab}\,.
 \end{align}
 When we define
 \begin{align}
     X_{\l}^{R}=\sum_{a,b}c_{ab}X_{ab}
 \end{align}
 to obtain the desired result~\eqref{eq:Schur-decomp}.

\bibliographystyle{JHEP}
\bibliography{Refs}


\end{document}